\documentclass[12pt,preprint]{aastex}

\newcommand{\msun}{{~{\rm M}_\odot}}

\newcommand{\etal}{{et~al.}~}

\shorttitle{}
\shortauthors{Hicks et al.}

\begin{document}

\title{A Multiwavelength Analysis of the Strong Lensing Cluster RCS 022434-0002.5 at $z=0.778$}

\author{A.K. Hicks\email{ahicks@alum.mit.edu}
\affil{Department of Astronomy, University of Virginia, P.O. Box 400325,
Charlottesville, VA 22904}}

\author{E. Ellingson\email{elling@casa.colorado.edu} 
\affil{Center for Astrophysics and Space Astronomy, University of Colorado at Boulder, Campus Box 389, Boulder, CO 80309}}

\author{H. Hoekstra\email{hoekstra@uvic.ca}}
\affil{Department of Physics \& Astronomy, University of Victoria, Elliott Building, 3800 Finnerty Rd, Victoria, BC, V8P 5C2}

\author{M. Gladders\email{gladders@uchicago.edu}}
\affil{Department of Astonomy and Astrophysics, University of Chicago, 5640 S. Ellis Ave, Chicago, IL 60637, USA}

\author{H.K.C. Yee\email{hyee@astro.utoronto.ca}}
\affil{Department of Astronomy and Astrophysics, University of Toronto, 50 St. George St., Toronto, ON, M5S 3H4, Canada}

\author{M. Bautz\email{mwb@space.mit.edu}}
\affil{MIT Kavli Institute for Astrophysics and Space Research, 77 Massachusetts Ave., Cambridge, MA 02139, USA}

\author{D. Gilbank\email{gilbank@astro.utoronto.ca}}
\affil{Department of Astronomy and Astrophysics, University of Toronto, 50 St. George St., Toronto, ON, M5S 3H4, Canada}

\author{T. Webb\email{webb@physics.mcgill.ca}}
\affil{McGill University Department of Physics, Rutherford Physics Building, 3600 Rue University, Montreal, Quebec, Canada, H3A 2T8}

\and

\author{R. Ivison\email{rji@roe.ac.uk}}
\affil{Astronomy Technology Centre, Royal Observatory, Blackford Hill, Edinburgh EH9 3HJ}

\begin{abstract}


We present the results of two (101 ks total) {\it{Chandra}} observations of the $z=0.778$ optically selected lensing cluster RCS022434-0002.5, along with weak lensing and dynamical analyses of this object.  An X-ray spectrum extracted within $\rm{R}_{2500}$ ($362$ $h_{70}^{-1}$ kpc) results in an integrated cluster temperature of $5.1^{+0.9}_{-0.5}$ keV.  The surface brightness profile of RCS022434-0002.5 indicates the presence of a slight excess of emission in the core.  A hardness ratio image of this object reveals that this central emission is primarily produced by soft X-rays.  Further investigation yields a cluster cooling time of $3.3 \times 10^9$ years, which is less than half of the age of the universe at this redshift given the current $\Lambda$CDM cosmology.  A weak lensing analysis is performed using HST images, and our weak lensing mass estimate is found to be in good agreement with the X-ray determined mass of the cluster.  Spectroscopic analysis reveals that RCS022434-0002.5 has a velocity dispersion of $900\pm180$ $\rm{km}~{\rm{s}^{-1}}$, consistent with its X-ray temperature.  The core gas mass fraction of RCS022434-0002.5 is, however, found to be three times lower than expected universal values.  The radial distribution of X-ray point sources within $\rm{R}_{200}$ of this cluster peaks at $\sim 0.7 \rm{R}_{200}$, possibly indicating that the cluster potential is influencing AGN activity at that radius.  Correlations between X-ray and radio (VLA) point source positions are also examined.   

\end{abstract}


\section{Introduction \label{s:intro}} 

Clusters of galaxies which exhibit gravitational lensing have been actively sought after since their discovery in the mid-1980s~\citep{soucail,lynds}.  This interest is due to the numerous scientific opportunities that these objects provide.  Because they magnify sources which lie at very high redshifts, they allow us to investigate forming galaxies in the early epochs of the universe in great detail~\citep[e.g.,][]{ebbels,franx,pettini,swinbank}.  Gravitational lensing can also be used as an independent probe of the mass distribution of clusters~\citep[e.g.,][]{kneib,tyson,broadhurst}.  In addition, clusters with multiple lensed sources contain information on the underlying cosmology of the universe~\citep{golse}.  Until recently the cumulative sample of strong lensing clusters was relatively small and mainly restricted to objects at redshifts of $z<0.5$~\citep{wu,williams}.  

The Red-Sequence Cluster Survey~\citep[RCS;][]{gladders00,gladders05} has recently discovered eight strong lensing clusters at high redshift~\citep{gladders03}.  One of the most striking objects discovered via the RCS survey is RCS022434-0002.5 at $z=0.778$~\citep{gladders02}.  This cluster possesses a rich array of lensing arcs that originate from several background sources, one of which lies at a (spectroscopically confirmed) redshift of $z=4.88$, another at $z=3.66$~\citep{swinbank07}, and a radial arc was found to be at $z=1.050$~\citep{sand05}.  

RCS022434-0002.5 (hereafter RCS0224-0002) was observed twice with the Chandra X-ray Observatory, initially for 15 kiloseconds, and more recently for 89 kiloseconds, the deepest single X-ray follow-up observation of an RCS cluster to date.  Here we present our analysis of these observations.  From the data we derive a spatially resolved surface brightness profile, cluster temperature, and mass.  In addition, we compare high resolution X-ray flux and hardness images of the cluster core with observations in the optical and radio.  Weak lensing and dynamical analyses are performed and compared to X-ray results.  A point source analysis of this field is also presented.  

Unless otherwise noted, this paper assumes a cosmology of $\rm{H}_0=70~\rm{km}~\rm{s}^{-1}~\rm{Mpc}^{-1}$, $\Omega_{\rm{M}}=0.3$, and $\Omega_{\Lambda}=0.7$.  Using this cosmology, $1\arcsec = 7.42~h_{70}^{-1}~{\rm{kpc}}$ at the cluster redshift. In addition, all error bars represent 68\% confidence levels.

\section{$Chandra$ Observations and Data Reduction \label{s:obs3}}

RCS0224-0002 was observed by the Chandra X-ray Observatory on 15 November 2002 for 12,327 seconds, and 9 December 2004 for 89,202 seconds.  The aimpoints of the observations were located on the S3 chip of the Advanced CCD Imaging Spectrometer (ACIS) CCD array.  Both observations were conducted in VFAINT mode, and reprocessed accordingly.  Upon obtaining the data, aspect solutions were examined for irregularities and none were detected.  Charged particle background contamination was reduced by excising time intervals during which the background count rate exceeded the average background rate by more than 20\%.  Bad pixels were removed from the resulting event files, and they were also filtered on standard grades.  After cleaning, the available exposure times were 12,051 seconds and 88,973 seconds, providing a total exposure of 101,024 seconds.

Combined images, instrument maps and exposure maps were created in the 0.29-7.0 keV band using the CIAO 3.3.0.1 tools DMCOPY and MERGE\_ALL, binning the data by a factor of 2.  Data with energies above 7.0 keV and below 0.29 keV were excluded due to background contamination and uncertainties in the ACIS calibration, respectively.  Data between 0.29 keV and 0.6 keV were included despite questionable low energy calibration for the purpose of increasing the overall signal-to-noise ratio of the observation.

\begin{figure}
\centerline{\includegraphics[width=5in]{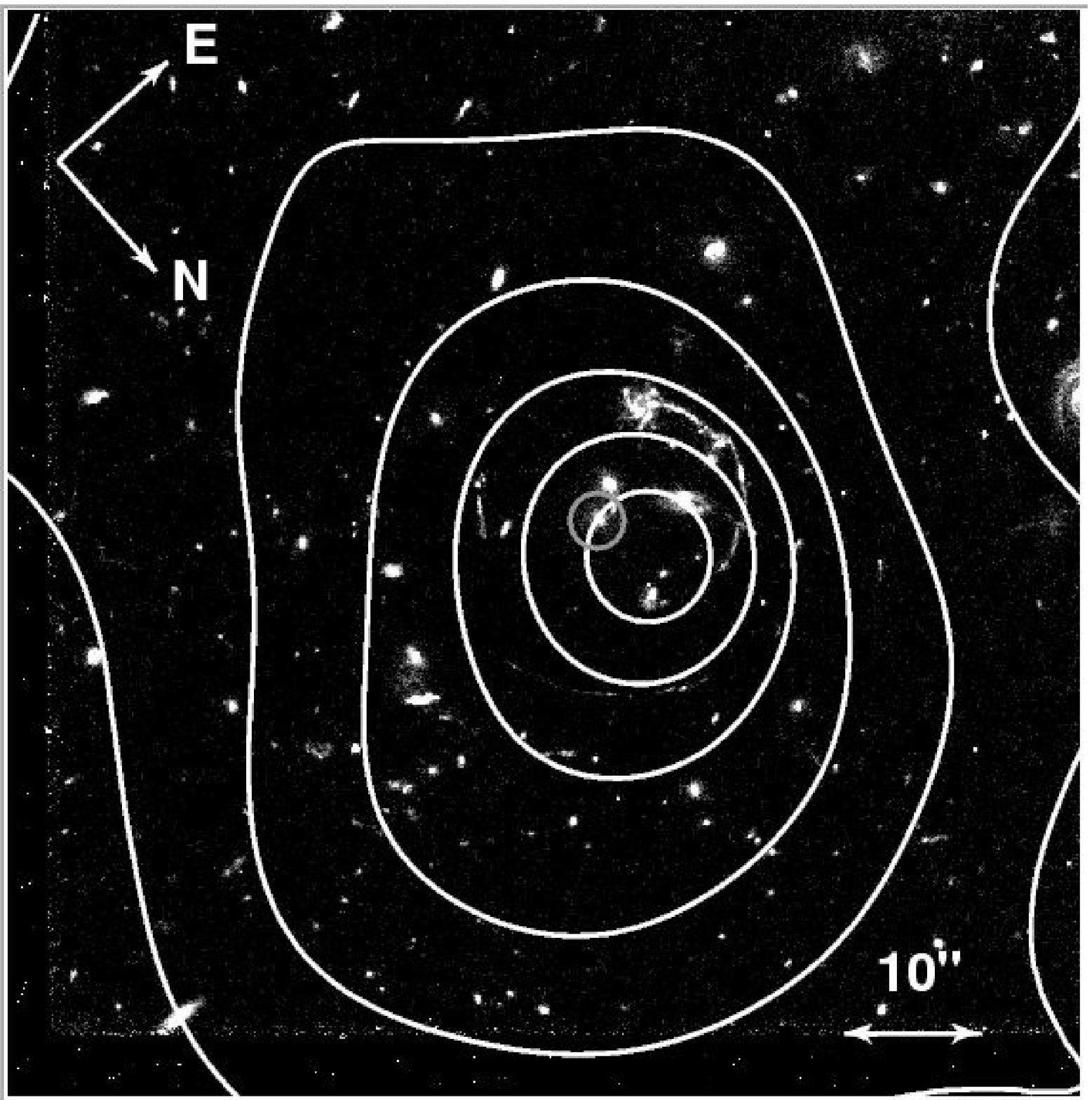}}
\caption{{\bf{RCS0224-0002 HST Overlay.}}  \label{fig1} logarithmic X-ray contours are overlaid on an 6600s HST image of RCS0224-0002 taken with the F606W filter.  X-ray countours were created from an adaptively smoothed Chandra 0.29-7.0 keV flux image, and have values between 1.25 and 0.3 $\times 10^{-8}~\rm{counts}~\rm{sec}^{-1}~\rm{cm}^{-2}~\rm{arcsec}^{-2}$.  The BCG is indicated by a small grey circle.  Note that the location of the X-ray centroid is displaced by $\sim4\arcsec$~from the center of the gravitationally-lensed arcs.}
\end{figure}

Flux images were created by dividing the image by the exposure map, and initial point source detection was performed by running the tools WTRANSFORM and WRECON on the 0.29-7.0 keV image.  These tools identify point sources using ``Mexican Hat'' wavelet functions of varying scales.  An adaptively smoothed flux image was created with CSMOOTH.  This smoothed image was used to determine the X-ray centroid of the cluster, which lies at an RA, Dec of 02:24:34.161, -00:02:26.44 (J2000).  The uncertainty associated with this position can be approximated by the smoothing scale applied to this central emission ($2.5\arcsec$).  Determined in this manner, the position of the X-ray centroid is $4.5\pm{2.5}\arcsec~(33.4\pm{18.6}$ $h_{70}^{-1}$ kpc) from the brightest cluster galaxy (BCG) of RCS0224-0002, falling within the range of typical offsets between BCGs and X-ray isophotal centers in X-ray selected samples~\citep{patel}.  Figure~\ref{fig1} shows an HST image of RCS0224-0002 with adaptively smoothed X-ray flux contours overlayed.

\section{Surface Brightness\label{s:beta}}

A radial surface brightness profile was computed from a combined, point source removed image and exposure map over the range
0.29-7.0 keV in $1\arcsec$ annular bins, then multiplied by a factor of $(1+z)^3$ to correct for cosmological dimming.  We find an excess of emission within the central $\sim 4 \arcsec$ of the cluster.  This central excess does not appear to be an AGN, since it is extended ($\sim4\times$ wider than the PSF) and no hard point source was detected at that location.   To test the validity of this central excess, a second surface brightness profile was computed in 2\arcsec~annular bins.

Both profiles were then fit with both single and double $\beta$ models.  A single $\beta$ model takes the form:

\begin{equation}
I(r) = I_B + I_0 \left( 1 + {r^2 \over r_c^2} \right)^{-3\beta+\frac{1}{2}},
\label{sb_eq}
\end{equation} 
where $I_B$ is a constant representing the surface brightness contribution of the background, $I_{0}$ is the normalization and $\rm{r}_{\rm{c}}$ is the core radius.


A double $\beta$ model is represented as:

 \begin{equation}
I(r) = I_B +
       I_{1} \left( 1 + {r^2 \over r_{1}^2} \right)^{-3\beta_1+\frac{1}{2}} + 
       I_{2} \left( 1 + {r^2 \over r_{2}^2} \right)^{-3\beta_2+\frac{1}{2}},
\label{sbdoub_eq}
\end{equation}

\noindent where each component has fit parameters $(I_n, r_n, \beta_n)$.

\begin{figure}
\centerline{\includegraphics[angle=90,width=5in]{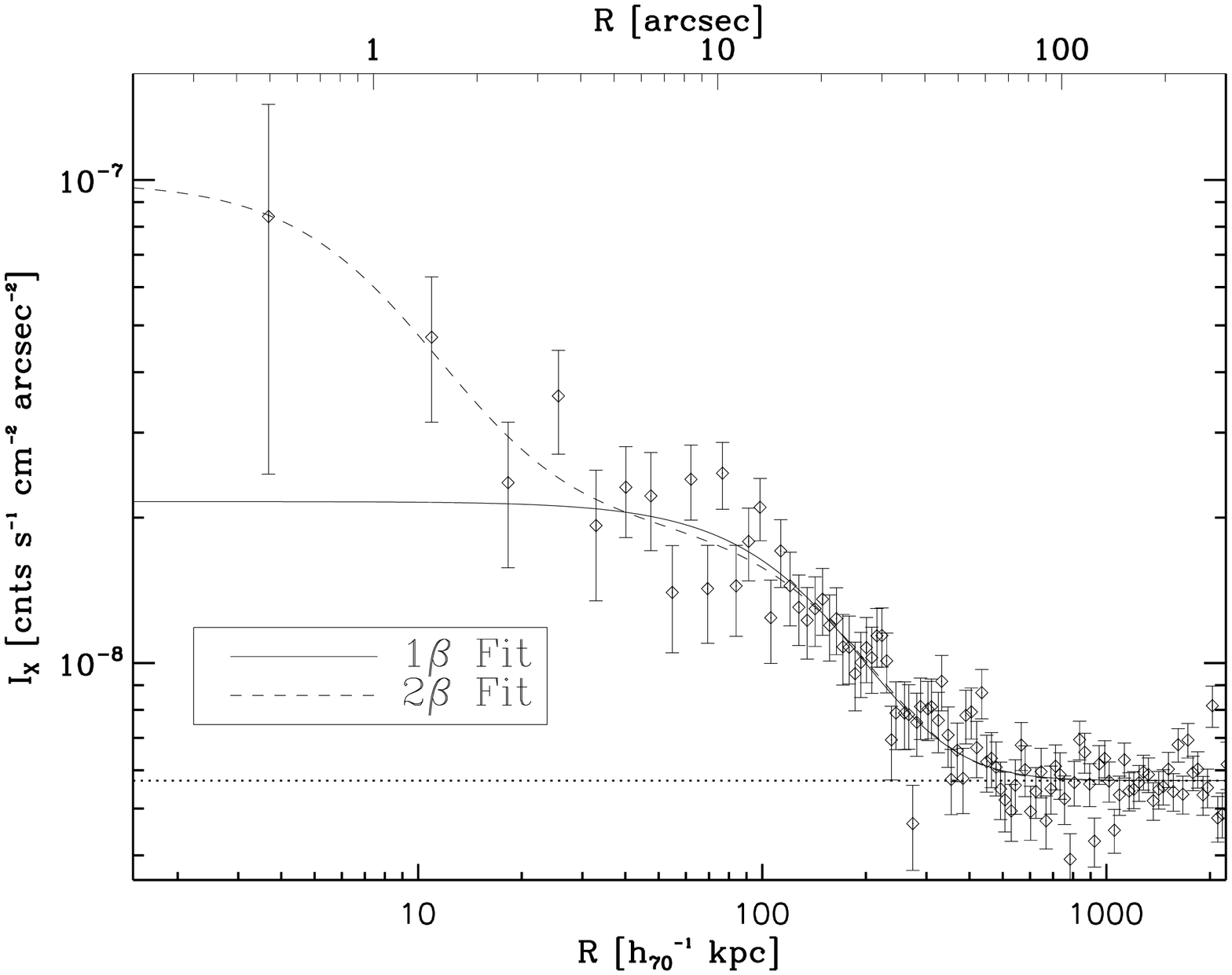}}
\centerline{\includegraphics[angle=90,width=5in]{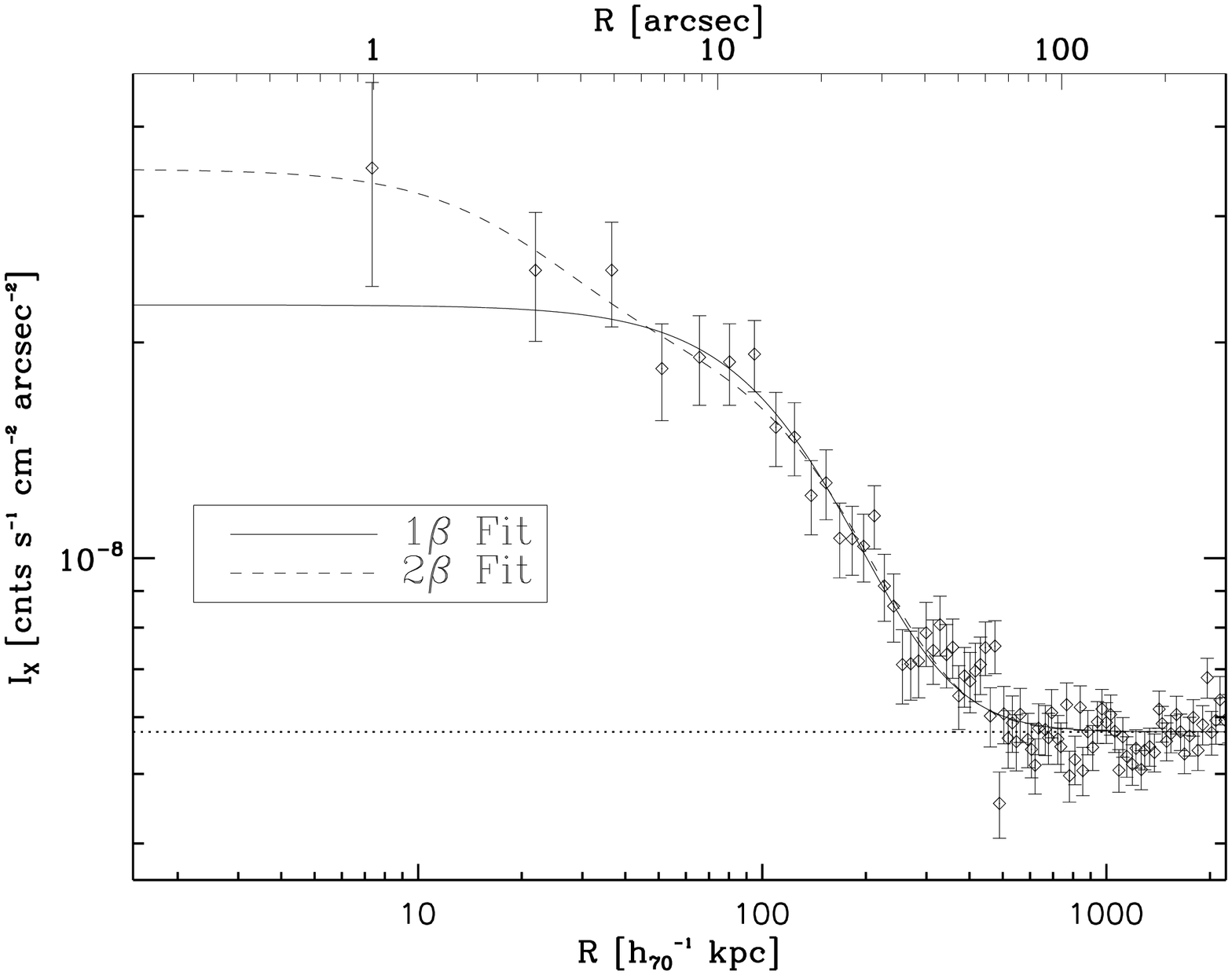}}
\caption{{\bf{RCS0224-0002 Surface Brightness Profiles.}}  \label{fig2} Radial surface brightness for the 0.29-7.0 keV band accumulated in $1\arcsec$ (top panel) and $2\arcsec$ (bottom panel) annuular bins.  Solid lines trace the best fitting single $\beta$ model fit and dashed lines illustrate the fits obtained by fitting this data with double $\beta$ models.  The dotted lines represent the best fitting background value for each profile.  Note the excess emission in the inner 20-30 $h_{70}^{-1}$ kpc of the cluster.}
\end{figure}

The central excess does persist in the more highly binned surface brightness profile, yet the reduced $\chi^2$s of the single vs. double $\beta$ model fits are almost identical in both cases, and there is no clear statistical preference between the two models.  $\beta$ model fits to both the 1\arcsec~and 2\arcsec~bin surface brightness profiles are shown in Figure~\ref{fig2} and details of all four fits are given in Table~\ref{table1}.  Unless otherwise noted, the single $\beta$ model fit to 2\arcsec~bin data is used in all subsequent analysis. 






\section{Spectral Analysis and $\rm{R}_{2500}$\label{s:spect3}}  

Using the centroid position indicated in Section~\ref{s:obs3}, a point source removed spectrum was extracted from each observation in a circular region with a $300~\rm{h}_{70}^{-1}~\rm{kpc}$ radius.  These spectra were then jointly analyzed in XSPEC~\citep{arnaud96}, using weighted response matrices (RMF) and effective area files (ARF) generated with the CIAO tool SPECEXTRACT and the Chandra calibration database version 3.2.2.  Backgrounds were extracted from the aimpoint chip in regions free of cluster emission.  The resulting spectra, one extracted from the short observation and one from the long observation, containing 183 and 1651 source counts respectively.  Both spectra were grouped to include at least 20 counts per bin.  

The spectra were jointly fit with a single temperature spectral model including foreground absorption.  The absorbing column was fixed at its measured value of $2.91 \times 10^{20} \rm{cm}^{-2}$~\citep{dickey90}, and the metal abundance was fixed at 0.3 solar~\citep{edge}.  Data with energies below 0.29 keV and above 7.0 keV were excluded from the fit.  

The results of these fits, combined with the single $\beta$ model parameters from Section~\ref{s:beta}, were then used to estimate the value of $\rm{R}_{2500}$ for this cluster.  This is accomplished by combining the equation for total gravitating mass

\begin{equation}
M_{tot}(<r) = -{{kT(r)r}\over{G \mu m_p}} \left({{\delta~\rm{ln}~\rho}\over{\delta~\rm{ln}~r}} + {{\delta~\rm{ln}~T} \over {\delta~\rm{ln}~r}} \right), 
\label{Mass_eq}
\end{equation}
 
\noindent
(where $\mu m_p$ is the mean mass per particle) with the definition of mass overdensity

\begin{equation}
M_{tot}(r_\Delta) = {{4}\over{3}}\pi\rho_c(z) r^3_\Delta \Delta ,
\end{equation}

\noindent
where $z$ is the cluster redshift, and $\Delta$ is the factor by which the density within $r_\Delta$ exceeds $\rho_c(z)$, the critical density at $z$. Here $\rho_c(z)$ is given by $\rho_c(z) = {3 {H_0^2 E(z)^2}/{8 \pi G}}$, where $E(z)=[\Omega_m(1+z)^3 + \Omega_{\Lambda}]^{1/2}$.  These equations are then combined with the density profile implied from the $\beta$ model (assuming hydrostatic equilibrium, spherical symmetry, and isothermality)

\begin{equation}
\rho_{gas}(r) = \rho_0 \left[1 + {{r^2}\over{r_c^2}}\right]^{-3\beta/2},
\label{dens_eq}
\end{equation}

\noindent
resulting in the equation

\begin{equation}
{{r_\Delta}\over{r_c}} = \sqrt{{\left[{{3\beta k T}\over{G \mu m_p (4/3) \pi \rho_c(z) r_c^2 \Delta}}\right]}-1},
\label{eq:ettori}
\end{equation}

\noindent
\citep{ettori00,ettori04b}.

\begin{figure}
\centerline{\includegraphics[angle=270,width=6in]{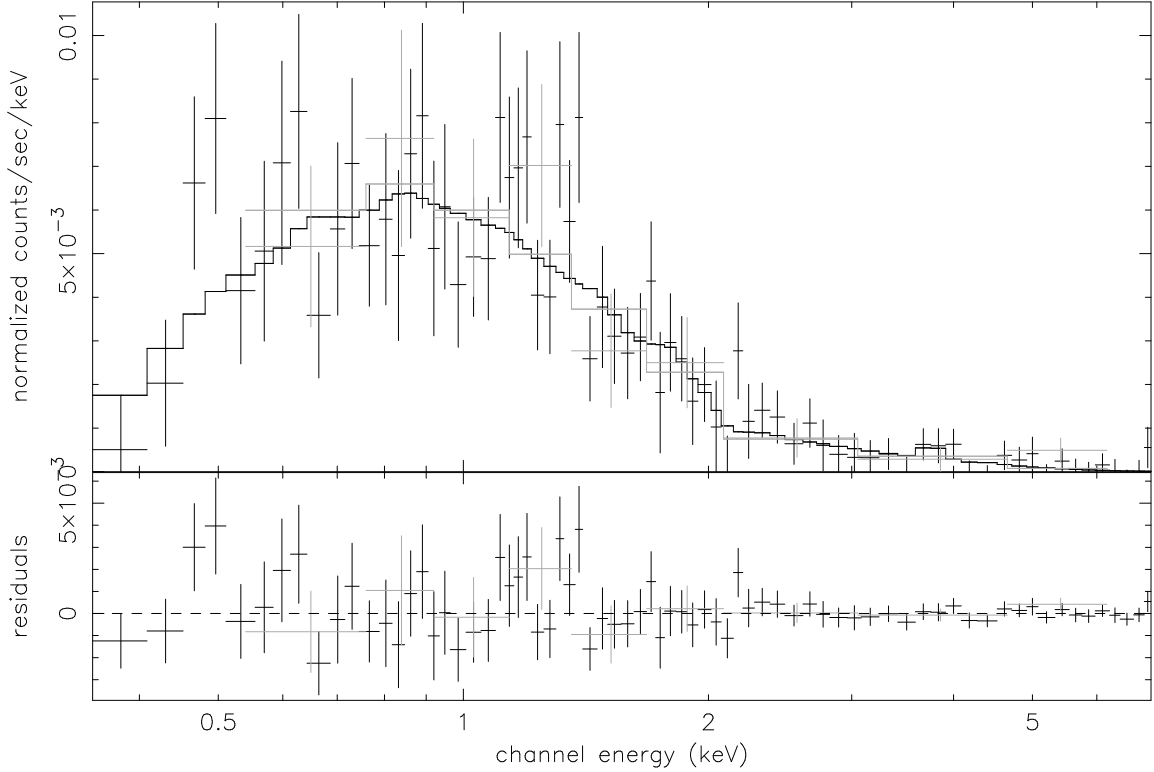}}
\caption{{\bf{RCS0224-0002 Core Spectra.}}  \label{fig3} XSPEC plot showing spectra fit with a single temperature spectral model.  The spectrum taken from the shorter observation is shown in grey.  Spectra were extracted from a circle centered on the emission peak with a radius of 362 $h_{70}^{-1}$ kpc ($\sim\rm{R}_{2500}$), and were grouped to include 20 counts per bin.  The fit resulted in an ambient cluster temperature of $5.1^{+0.9}_{-0.5}$ keV.}
\end{figure}

After the initial estimation of $R_{2500}$, additional spectra were extracted from within that radius, and spectral fitting was performed again.  This iterative process was continued until the fitted temperature and calculated value of $R_{2500}$ was consistent with the extraction radius.  The final estimate of $R_{2500}$ is $362^{+61}_{-46}$ $\rm{h}_{70}^{-1}$ kpc (48.8\arcsec), with an extrapolation to $R_{200}$ of $1.47^{+186}_{-141}$ $\rm{h}_{70}^{-1}$ Mpc.  The best fit temperature of the spectra extracted from within this region was $\rm{T}_{\rm{x}}=5.1^{+0.9}_{-0.5}$ keV, with a reduced $\chi^2$=0.75 for 77 degrees of freedom.  Using this spectrum, the unabsorbed bolometric luminosity of RCS0224-0002 within $R_{2500}$ is $L_X = 2.3^{+0.2}_{-0.3} \times 10^{44} \rm{erg}~{\rm{s}}^{-1}$.  Figure~\ref{fig3} is a plot of these spectra with the best fitting model overlaid on the data.

\section{Hardness Map\label{s:hard}}

Images, instrument maps and exposure maps were created for both a soft X-ray band (0.5-2.0 keV) and a hard band (2.0-8.0 keV) by the method described in Section~\ref{s:obs3}.  These files were then used to create flux images in the two bands.  

Using the smoothing scale produced in the analysis of the entire energy band, smoothed flux images were created in each of the two bands.  The hard band image was subtracted from the soft band image, and the result of that calculation was divided by the smoothed flux image of the entire energy band (i.e., ${\rm{I}_{\rm{S}}-\rm{I}_{\rm{H}}}\over{\rm{I}_{\rm{tot}}}$).  Contours of the resulting hardness ratio image are overlayed on a VLA 1.4 GHz radio image in Figure~\ref{fig4}.  Soft emission is concentrated in a region containing both the X-ray centroid and the BCG, a radio galaxy to the southwest of the X-ray centroid (Figure~\ref{fig4}). There is no clear emission peak that coincides with the BCG, a radio galaxy at $z=0.7773$. 


\section{Mass Determinations}
 
\subsection{Gas Density and Cooling Time}

The gas mass density distribution of a cluster of galaxies whose
surface brightness profile is well described by a $\beta$ model can be
shown to follow Equation~\ref{dens_eq}. This relationship is dependent upon the assumptions of hydrostatic 
equilibrium, spherical symmetry, and gas isothermality.  The central
density in this equation can be determined by an expression relating
the observable cluster X-ray luminosity to gas density.  This
calculation requires the experimental determination of both gas
temperature and emission measure, which can be performed in XSPEC.    

With the results of spectral analysis (Section~\ref{s:spect3}) and using the best fitting surface brightness parameters for the inner luminosity peak (Section~\ref{s:beta}), a central gas density of $7.6\pm{0.09}\times 10^{-3}~\rm{cm}^{-3}$ is obtained for RCS0224-0002.
  
Recall that the characteristic time that it takes a plasma to cool isochorically
through an increment of temperature $\delta$T can be written

\begin{equation}
  \delta t_{cool} = \frac{3}{2}~\frac{k}{n~\Lambda(T)}~\delta T , 
\label{cool}
\end{equation}

\noindent \citep{sarazin88}, where $n$ is the electron density, $\Lambda$(T) is the
total emissivity of the plasma (the cooling function), and k is
Boltzmann's constant.  In the case of isobaric cooling, 3/2 is
replaced by 5/2. 

Using this formula combined with the cooling function appropriate to a 0.3 solar abundance plasma, the cooling time of RCS0224-0002 is $3.3\times 10^9$ years.  The age of the universe in our adopted cosmology at $z=0.778$ is $6.8\times10^9$ years~\citep[e.g.,][]{thomas}.




Therefore the cooling time of the innermost region of RCS0224-0002 is less than the age of the universe at $z=0.778$.  This, combined with the luminosity excess in the central few arcseconds, as well as the existence of a soft X-ray peak at the same location, indicates that this cluster may contain a proto-cooling core.  The observation does not contain enough source photons to obtain spatially resolved temperature information for this cluster, therefore longer exposures will be required to completely characterize the central emission.   

At least one other study has detected a high redshift cluster with the signatures of a small cooling core.  \citet{grego04} in their analysis of a Chandra observation of MS1137.5+6625 (at $z=0.783$) detect a small central surface brightness excess, and an excess of soft emission in the core of this cluster.  They also calculate a central cooling time that is shorter than the age of the universe at that epoch.

\subsection{X-ray Mass Estimate}

Using the results of the $\beta$ model fits (Section~\ref{s:beta}) and spectral fit (Section~\ref{s:spect3}), along with Equation~\ref{dens_eq} and the equation of hydrostatic equilibrium (Equation~\ref{Mass_eq}), it can be shown that

\begin{equation}
M_{tot}(<r) = {{3\beta}\over{G}}~{{k T r}\over{\mu m_p}}~{{{(r/r_c)}^2}\over{{1+{{(r/r_c)}^2}}}}. 
\label{Mass_eq2}
\end{equation}

\noindent Using this equation (with $\mu=0.62$), gas mass and total mass were calculated out to $\rm{R}_{2500}$ ($\sim362$ $h_{70}^{-1}$ kpc) and $\rm{R}_{200}$ ($\sim 1.47$ $h_{70}^{-1}$ Mpc) for RCS0224-0002.  One sigma errors on central density, gas mass, total mass, and gas mass fraction were determined through statistically analyzing a random sampling ($N=10,000$) of values drawn from the 68\% confidence intervals of relevant parameters, assuming that their errors are uncorrelated.  Mass determinations resulted in $\rm{M}_{\rm{2500,tot}}=1.6\pm{0.2} \times 10^{14}~h_{70}^{-1} \msun$, $\rm{M}_{\rm{2500,gas}}=0.05\pm{0.005} \times 10^{14}~h_{70}^{-5/2} \msun$, $\rm{M}_{\rm{200,tot}}=8.3\pm{0.8} \times 10^{14}~h_{70}^{-1} \msun$ and $\rm{M}_{\rm{200,gas}}=0.24\pm{0.02} \times 10^{14}~h_{70}^{-5/2} \msun$.  The mass-temperature relation of~\citet{vikhlinin06} predicts $\rm{M}_{\rm{2500,tot}}=1.2\pm{0.05} \times 10^{14}~h_{70}^{-1} \msun$ for $z=0.778$ and $T_X=5.1$ keV, which is within $2\sigma$ (25\%) of our observed value. 

The core gas mass fraction of RCS0224-0002 is $f_{g,2500}=0.031\pm{0.005}~h_{70}^{-3/2}$.  This value is significantly lower than that determined via the same method for a sample of 14 medium redshift X-ray selected clusters~\citep{hicks06} and is three to four times lower than both theoretically and observationally expected ``universal'' values~\citep[][]{evrard,allen04}.  This is a phenomenon that is also seen in groups, poor clusters, and other high redshift clusters~\citep{dellantonio,sanderson,sadat}, as well as high redshift cluster simulations~\citep{ettori04b} and additional RCS clusters~\citep[in prep]{hicksrcs}.


\subsection{Weak Lensing Analysis\label{s:lens}}

The central region of RCS0224-0002 was observed in the $F606W$ and $F814W$
filter with the WFPC2 camera onboard the Hubble Space Telescope, with the center of the cluster located on
WFC3. The observations consist of 2 orbits in each filter, yielding
total integration times of 6600s in $F606W$ and $F814W$. 

The exposures in each passband were split into two sets (by the orbit
the data were taken), resulting in two images. These two images were
designed to be offset by 5.5 pixels, to allow the construction of an
interlaced image \citep[e.g.,][]{vandokkum99,hoekstra00}, which
has a sampling that is a factor $\sqrt{2}$ better than the original
WFPC2 images. This same approach was used by \citet{hoekstra00}
in their study of MS1054-03 and we refer to this paper for more
details. For the analysis presented here we omitted data from the
Planetary Camera because of the the brighter isophotal limit.

The next step is to use the interlaced images for our weak lensing
analysis. The analysis technique is based on the one developed by
\citet{kaiser95} with a number of modifications described in
\citet{hoekstra98} and \citet{hoekstra00}.  We note that this
implementation has been tested extensively \citep[e.g.,][]{hoekstra98, heymans06} and has been shown to be accurate at
the few percent level. We follow the procedure outlined in \citet[][2000]{hoekstra98} which results in a catalog of ellipticities for the
faint galaxies that we use in our lensing analysis. These shapes have
been corrected for PSF anisotropy and the size of the PSF.

The measurement of the shape of an individual galaxy provides only
a noisy estimate of the weak gravitational lensing signal. We therefore
average the measurements of a large number of galaxies as a function
of distance to the cluster center. As discussed in \citet{hoekstra00}
we weight each object with the inverse square of the uncertainty in
the distortion, which includes the contributions of the intrinsic
ellipticities and the shot noise in the shape measurement.

Figure~\ref{fig5} shows the resulting average tangential distortion as a
function of distance from the cluster center, which was taken to be
the location of the brightest cluster galaxy. We selected 462 galaxies
with $22.5<F814W<25.5$ and $F606W-F814W<1.5$ (i.e., bluer than the
cluster early types) as source galaxies. We detect a significant
lensing signal. The lower panel of Figure~\ref{fig5} is the signal, $g_X$,
when the phase of the distortion is increased by $\pi/2$. If the signal
observed in the upper panel is due to gravitational lensing, $g_X$ should
vanish, as is observed. The solid line is the best fit singular
isothermal sphere model which yields an Einstein radius of $4.7\pm0.9$ arcseconds.

To relate the observed lensing signal to an estimate of the mass
requires an estimate of the mean source redshift distribution. As in
\citet{hoekstra00}, we use the results from \citet{fernandez99} which are based on the Hubble Deep Fields. This allows us
to determine an equivalent line of sight velocity dispersion using the best fit
isothermal sphere model. We obtain a value of $\sigma=777^{+69}_{-76}$
$\rm{km}~\rm{s}^{-1}$.

Alternatively we can fit an NFW model to the measurements. We assume
that the density profile is only a function of the virial mass and use
the relation between concentration and mass from \citet{bullock01}. This yields a virial mass of $5.4^{+4.1}_{-2.7}\times
10^{14} \msun$ and a corresponding $M_{2500}=1.3^{+1.0}_{-0.6}\times
10^{14}\msun$ (and $r_{2500}=375$ kpc). This result is in excellent
agreement with both the dynamical and X-ray analysis results.

As shown in \citet{hoekstra02} mass estimates based on single HST
pointings tend to be biased low, because of substructure in the
cluster center. However, the distribution of galaxies in RCS0224-002
is relatively regular and does not show significant substructure.
Furthermore, because of its high redshift, the lensing signal can be
measured sufficiently far out to be less affected by substructure.

\subsection{Dynamical Analysis}

Optical multi-object spectroscopic observations of galaxies in this field were performed by several telescopes and instruments; dates and observational parameters are listed in Table~\ref{table2}. Both galaxies and gravitational arc images were targeted, over a field of view of 5 arcminutes. Most cluster galaxies had magnitudes of R$\sim$ 20-22, though a few galaxies as faint as R$\sim 24$ were identified via their [OII] emission lines.   Several slits targeting cluster arcs were tilted or curved \citep[see][]{gladders03}, but these were not used for cluster galaxy redshifts. Overall, 39 galaxy redshifts at $0.70 < z < 0.85$ were  measured by the wavelength of individual lines or via cross-correlation against galaxy spectra,  with a range of spectral types. We find no significant systematic differences in the velocities measured by the different telescopes and instruments, either in velocity centroid, or in the 3 objects whose redshifts were measured accurately by more than one telescope.

 Figure~\ref{fig6} shows a histogram of galaxy velocities. There are two secondary peaks close to the cluster, one at z$\sim 0.750$ and another at z$\sim 0.805$. Both of these comprise galaxies spread over the field of view and are not associated with an easily identified isolated clump. There are substantial gaps between them and the cluster peak at z=0.778, and hence we do not identify these galaxies as cluster members. 

 
Correcting in quadrature for an estimated uncertainty of 100 $\rm{km}~\rm{s}^{-1}$ for individual velocity measurements, we find that the cluster is located at a mean redshift of 0.7777 and has a rest-frame velocity dispersion of 900 $\pm$ 180 $\rm{km}~\rm{s}^{-1}$, based on 24 cluster members. This estimate is very close to that expected from a standard $\sigma$-T$_X$ relationship \citep{voit03,xue00}, where an X-ray temperature of 5.1 keV predicts a velocity dispersion of 872 $\rm{km}~\rm{s}^{-1}$.  Our velocity dispersion measurement is also consistent with both weak lensing analysis (\ref{s:lens}) and the value of 920 $\rm{km}~\rm{s}^{-1}$ reported by ~\citet{swinbank07} based on their strong lensing analysis of this cluster.


\section{Point Sources}

\subsection{X-ray Source Population}

Studies of the AGN population in clusters of galaxies are an important probe of galaxy evolution in these environments~\citep{martini}.  While AGN can be difficult to detect at optical wavelengths due to obscuration, they are much more easily seen in the X-ray, where they comprise the dominant contribution to the hard X-ray point source population.  In this section we perform an X-ray point source analysis of the Chandra observation of RCS0224-0002.  

Using the hard and soft band images and exposure maps referred to in Section~\ref{s:hard}, point sources in these bands were identified using the CIAO tools WTRANSFORM and WRECON.  Only sources with a detection significance above $3\sigma$ were retained for further analysis.  In the case of faint sources, our numbers may represent a lower limit, due to both off-axis degradation of the PSF~\citep[][]{kim}, and the possible masking of very faint sources by diffuse X-ray emission in the cluster center. 

Using PIMMS (Portable Interactive Multi-Mission Simulator), assuming an absorbed powerlaw spectrum with a hydrogen column density fixed at the galactic value and a powerlaw index of 1.6, X-ray fluxes were calculated for each source after correcting for exposure variations across the chips.  Log$N$-log$S$ information was then calculated for each energy band, inclusive of all sources within $\rm{R}_{200}$.  These data were then compared to the log$N$-log$S$ background values of~\citet{moretti}.  The results of these comparisons with $1 \sigma$ error bars~\citep{gehrels} are shown in Figure~\ref{fig7}.  The data are generally consistent with the measured X-ray background, however the background data do not extend to very low flux sources, and there may be a very slight excess of low-flux soft X-ray sources within $\rm{R}_{200}$.  ``Cosmic variance'', however, can become significant at low-flux as well, showing variations in source counts as high as $3.9\sigma$ for $2.0-8.0$ keV sources with fluxes below $<1\times10^{-15} \rm{erg}~\rm{cm}^{-2}~\rm{s}^{-1}$~\citep{brandt}

A list of point sources within $\rm{R}_{200}$ is given in Tables~\ref{table3} and Table~\ref{table4} for the 0.5-2.0 keV and 2.0-8.0 keV bands, respectively.  These tables include both flux and luminosity information (also achieved with PIMMS, assuming a redshift of $z=0.778$) for each source.  


To compare the distribution of sources in the RCS0224-0002 field to those reported in~\citet[][based on a statistical study of $\sim50$ clusters]{ruderman}, radial histograms containing the number of sources per square megaparsec (assuming a redshift of $z=0.778$) were created from data in each energy band (Figure~\ref{fig8}).  These plots indicate a significant excess of point sources within $\rm{R}_{200}$, particularly in the $0.6-0.8~\rm{R}_{200}$ bins.  The~\citet{ruderman} study was conducted in the 0.5-2.0 keV energy band, and our soft band data are roughly consistent with their results. 


\subsection{Radio Comparisons}

RCS0224-0002 was observed with the National Radio Astronomy Observatory's (NRAO) Very Large Array (VLA)\footnote{The National Radio Astronomy Observatory is a facility of the National Science Foundation operated under cooperative agreement by Associated Universities, Inc.} at 1.4 GHz for a total of 15 hours over three days~\citep{webb}.  Using this data, an investigation of the coincidence of X-ray point sources with 1.4 GHz sources was performed.  Figure~\ref{fig9} consists of the 1.4 GHz radio image of RCS0224-0002 with 0.2-5.0 and 2.0-8.0 X-ray point source regions overlaid.  Obvious correlations are noted in Tables~\ref{table3} and~\ref{table4}, and can be seen in Figure~\ref{fig9}.  None of the detected X-ray point sources (in either energy band) is coincident with the cluster BCG.

RCS0224-0002 was also observed at both 450$\mu$m and 850$\mu$m with the Submillimeter Common-User Bolometer Array~\citep[SCUBA]{holland} in 2001 and 2002.  The results of these observations are reported in~\citet{webb}, and have resulted in five submillimeter source detections in the RCS0224-0002 field.  Only one of the five sources is coincident with an X-ray detected point source, smm-RCS0224.2 (source 6 in Table~\ref{table3} and source 3 in Table~\ref{table4}).

\section{Discussion and Summary\label{sum}}

This paper describes the results of the most in depth X-ray analysis of an optically selected high redshift RCS cluster to date, the 101 ks of Chandra observations of the lensing cluster RCS0224-0002 at $z=0.778$.  Imaging analysis reveals that the peak of X-ray emission is slightly displaced ($4.5\pm{2.5}\arcsec$) from the center of the lensing arcs.  This displacement falls well within the common range of X-ray centroid offsets~\citep{patel}.

Single temperature spectral fitting of the X-ray emission within a 362 $h_{70}^{-1}$ kpc ($\rm{R}_{2500}$) radius indicates that this cluster has an ambient temperature of $5.1^{+0.9}_{-0.5}$ keV.  The surface brightness profile of RCS0224-0002 exhibits a small central peak of emission which can be modeled via the inclusion of a second $\beta$ component.  The $\beta$ model used in our analysis has a core radius of $213^{+14}_{-13}$ $h_{70}^{-1}$ kpc and $\beta$ value of $0.88^{+0.07}_{-0.04}$.  Using the results of both temperature and surface brightness fitting, $\rm{R}_{200}$ is determined to be $1.47^{+0.1}_{-0.07}$ $h_{70}^{-1}$ Mpc for this cluster.

\begin{figure}
\centerline{\includegraphics[width=3in]{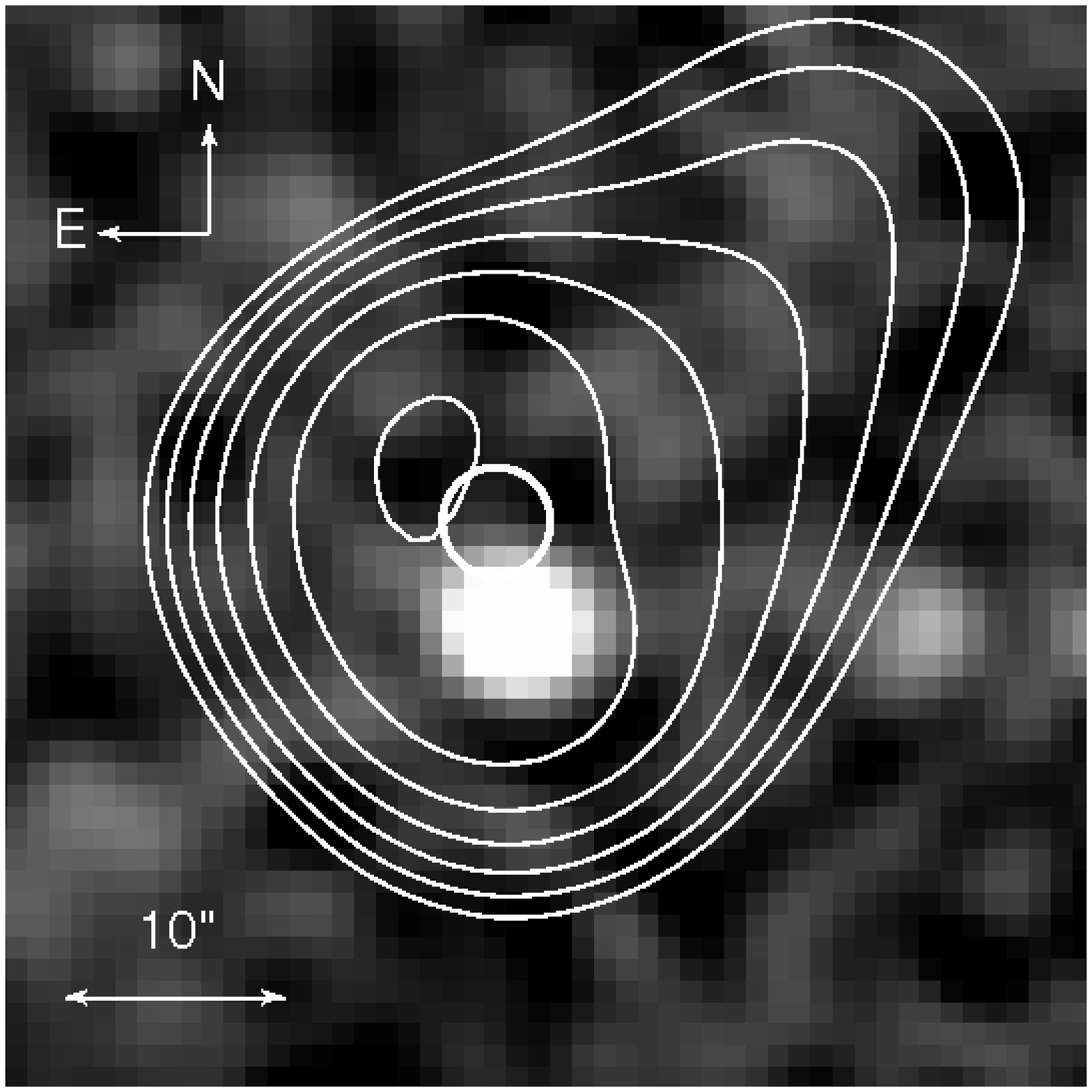}
\includegraphics[width=3in]{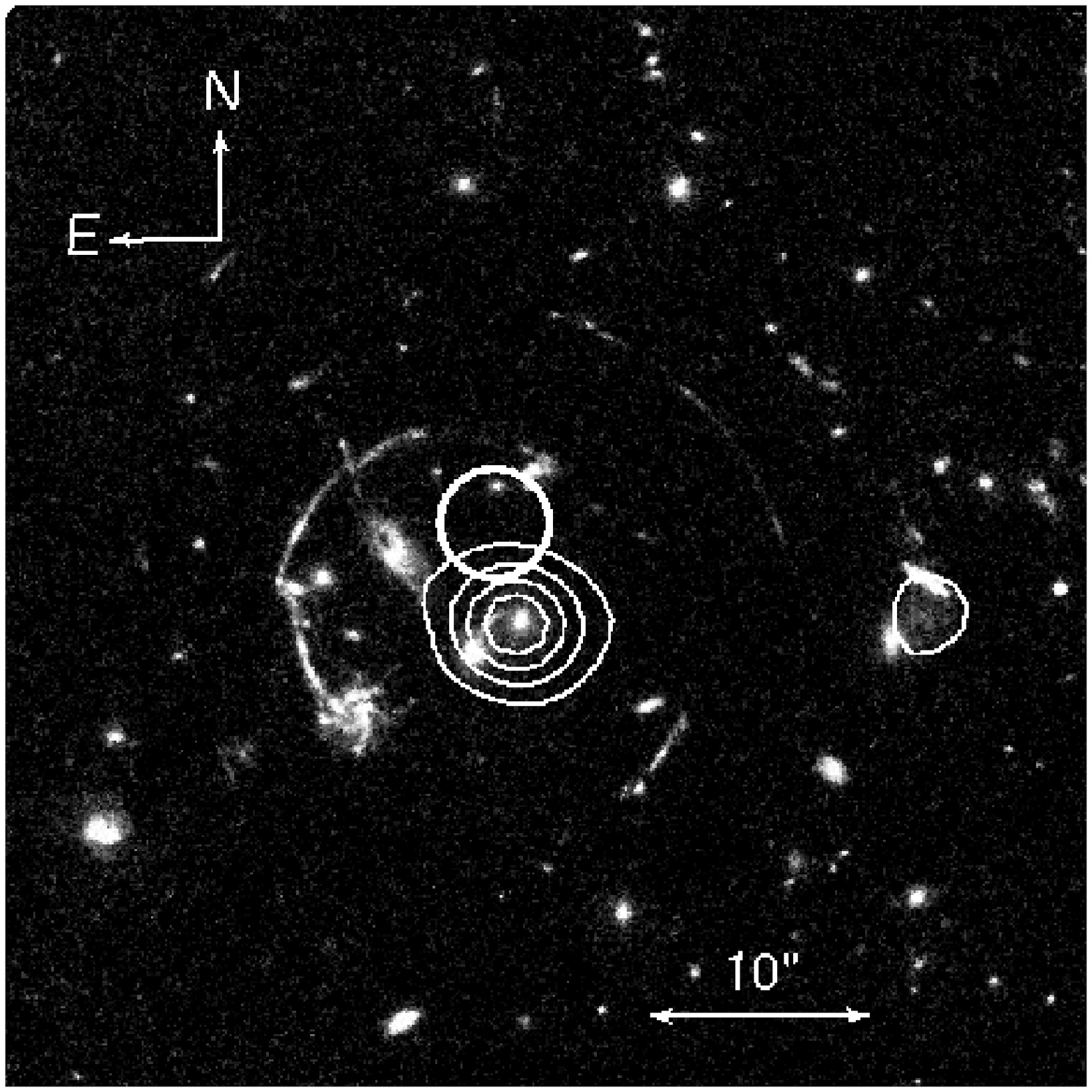}}
\caption{{\bf{Hardness Map and Radio Comparisons.}}  \label{fig4} {\bf{Left panel}} -  A hardness map was constructed for RCS 0224-0002 by dividing smoothed flux images in the hard (2.0-8.0 keV) and soft (0.5-2.0 keV) bands by a smoothed flux image of the total band in the manner ${\rm{I}_{\rm{S}}}-{\rm{I}_{\rm{H}}}\over{\rm{I}_{\rm{tot}}}$.  Linearly spaced contours were created from the hardness map and are overlaid on a 1.4 GHz VLA flux image (gray scale) of RCS0224-0002.  Soft emission peaks in a region containing both the X-ray centroid (shown as a thick 2.5\arcsec~radius circle) and a 1.4 GHz radio source.  {\bf{Right Panel}} - HST image of RCS0224-0002 (Figure~\ref{fig1}) with radio contours overlaid.  1.4 GHz VLA contours from the flux image shown in the left panel clearly indicate the galaxy which is responsible for the radio emission.  The position of the X-ray centroid is again represented as a bold 2.5\arcsec~radius circle.}
\end{figure}

\begin{figure}
\centerline{\includegraphics[width=6in]{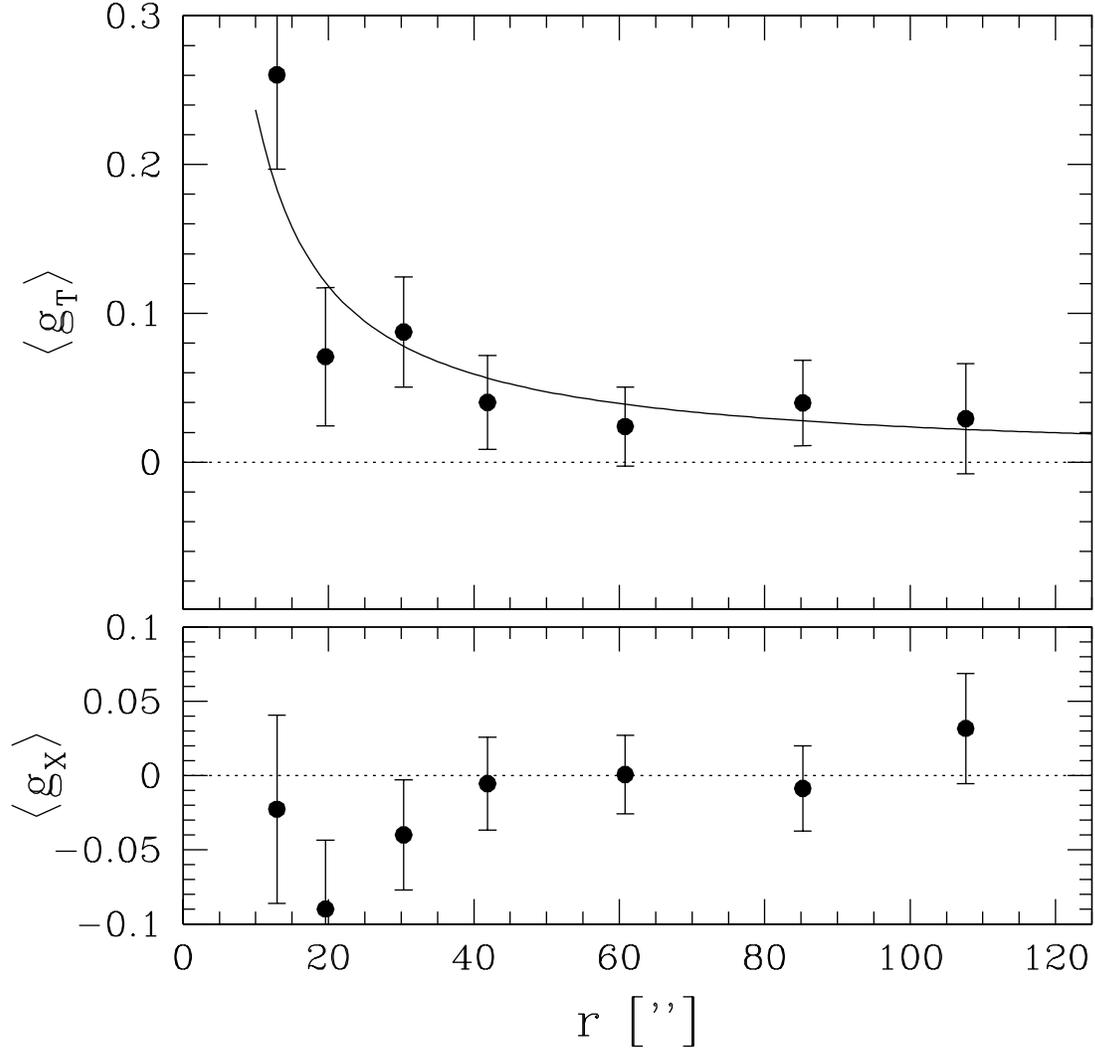}}
\caption{{\bf{Average Tangential Distortion}}  Average tangential distortion is plotted as a function of distance from the brightest cluster galaxy, using 462 source galaxies with $22.5<F814W<25.5$ and $F606W-F814W<1.5$ (i.e., bluer than the
cluster early types). We detect a significant
lensing signal.  The lower panel shows the signal
when the phase of the distortion is increased by $\pi/2$. If the signal
observed in the upper panel is due to gravitational lensing, $g_X$ should
vanish, as is observed. The solid line is the best fit singular
isothermal sphere model which yields an Einstein radius of $4.7\pm0.9$ arcseconds.  \label{fig5}}
\end{figure}

\begin{figure}
\centerline{\includegraphics[width=6in]{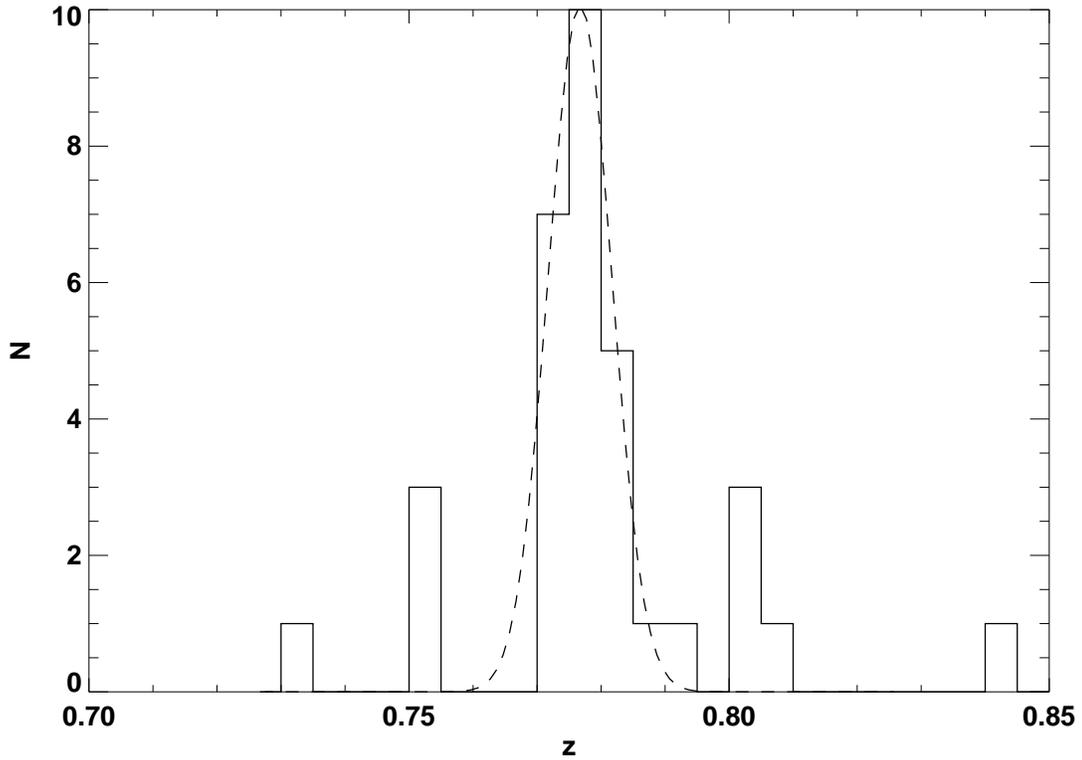}}
\caption{{\bf{Histogram of Galaxy Velocities.}} ~Histogram constructed with 39 galaxies, indicating a mean redshift of $z=0.778$ for RCS0224-0002.  The overplotted dashed line shows a Gaussian of dispersion 900 $\rm{km}~\rm{s}^{-1}$ rest-frame.  Note that the two secondary peaks at $z\sim0.750$ and $z\sim0.805$ are visibly separated from the cluster peak and are not spatially concentrated, so we do not associate these with the cluster nor with coherent structures.\label{fig6}}
\end{figure}

\begin{figure}
\centerline{\includegraphics[angle=90,width=5in]{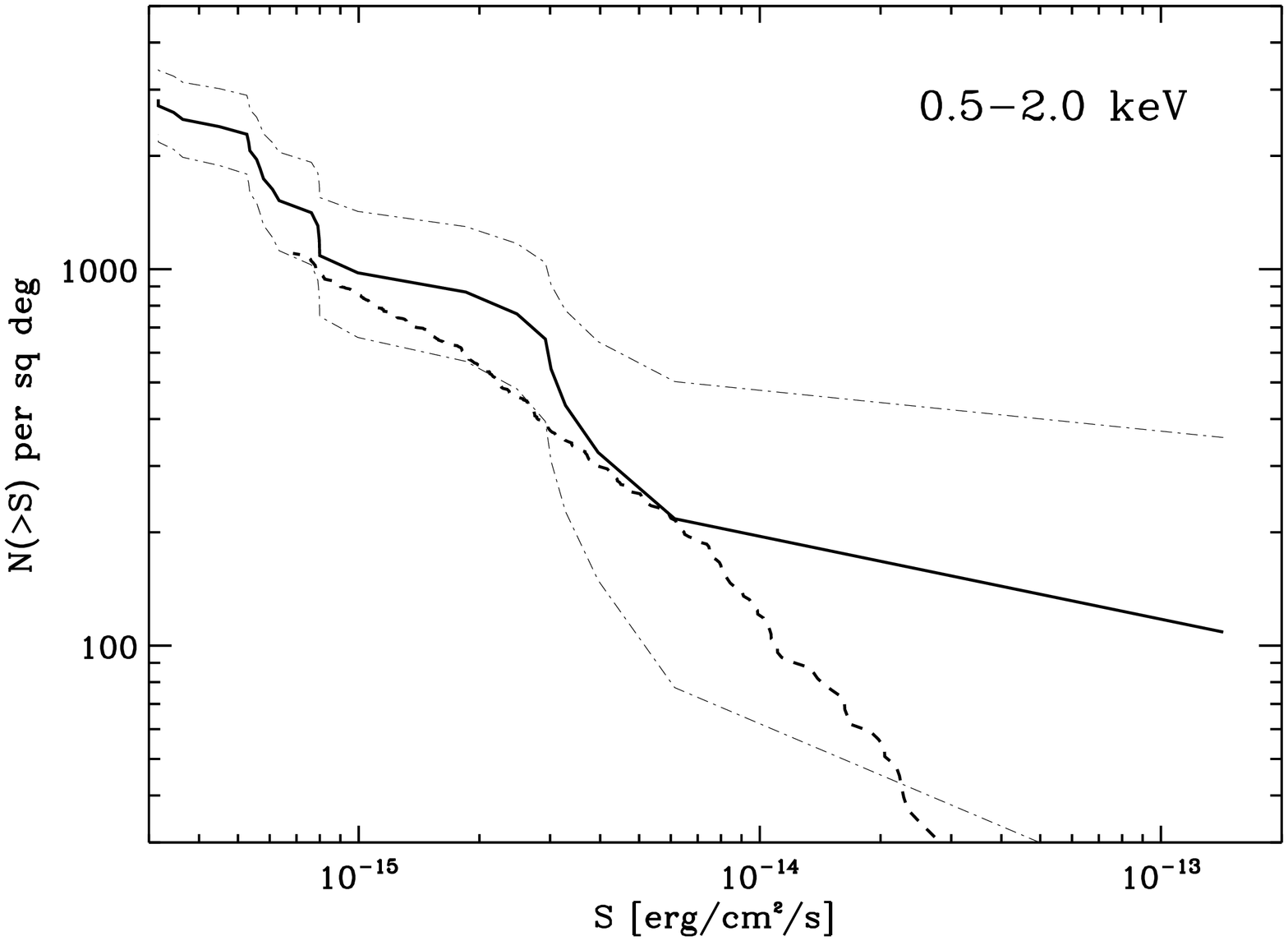}}
\centerline{\includegraphics[angle=90,width=5in]{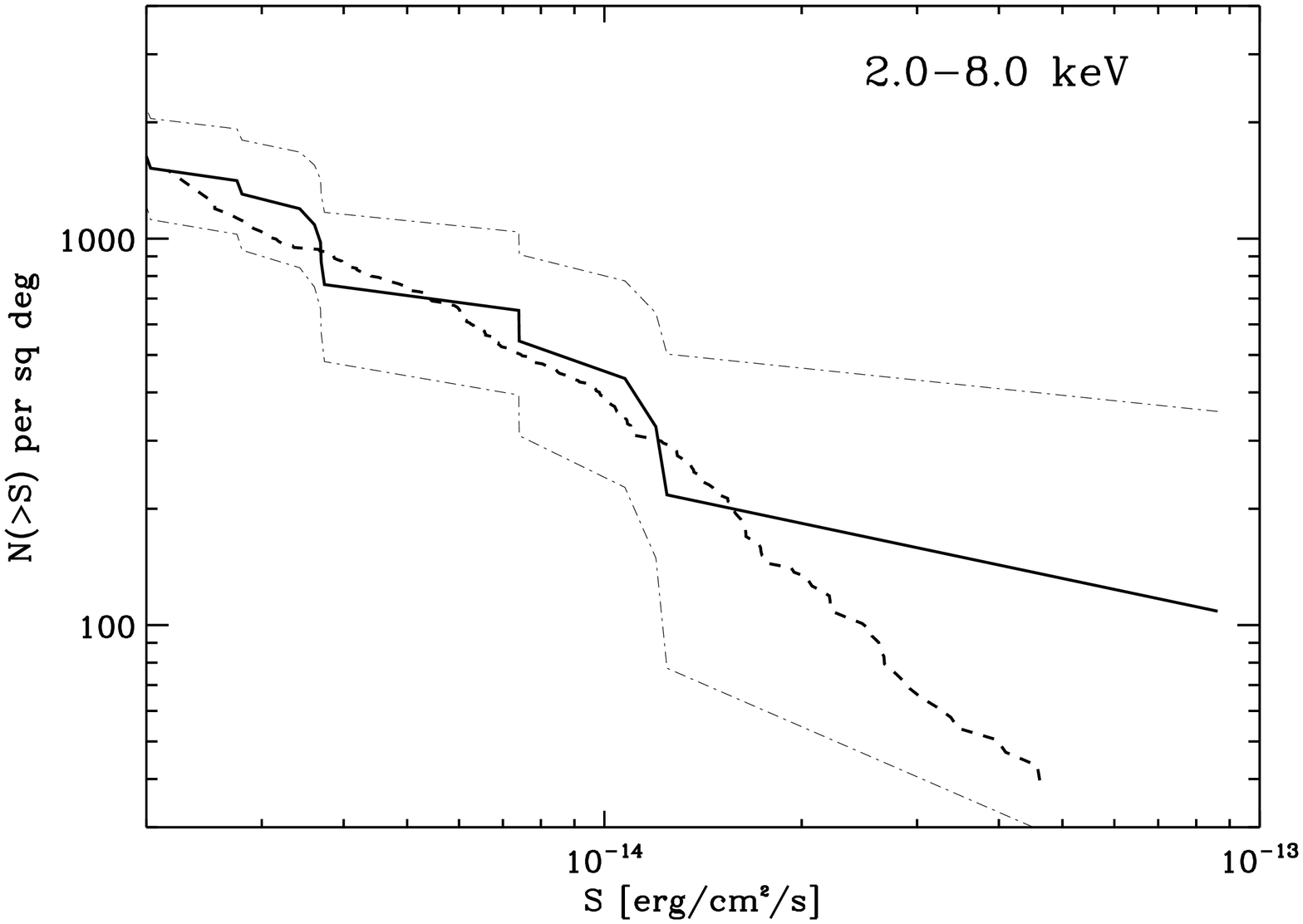}}
\caption{{\bf{X-ray Point Sources.}}  LogN-LogS plot of point sources within $\rm{R}_{200}$ (1.47 $h_{70}^{-1}$ Mpc) in soft (0.5-2.0 keV) and hard (2-8 keV) X-ray bands.  The dashed lines represent the X-ray background in these energy bands~\citep{moretti}.  The solid line in each plot indicates the LogN-LogS distribution of detected sources.  Dot dash lines indicate $1\sigma$ confidence intervals.  A slight excess of low flux point sources may be present in the soft energy band.\label{fig7}}
\end{figure}

\begin{figure}
\centerline{\includegraphics[angle=90,width=5in]{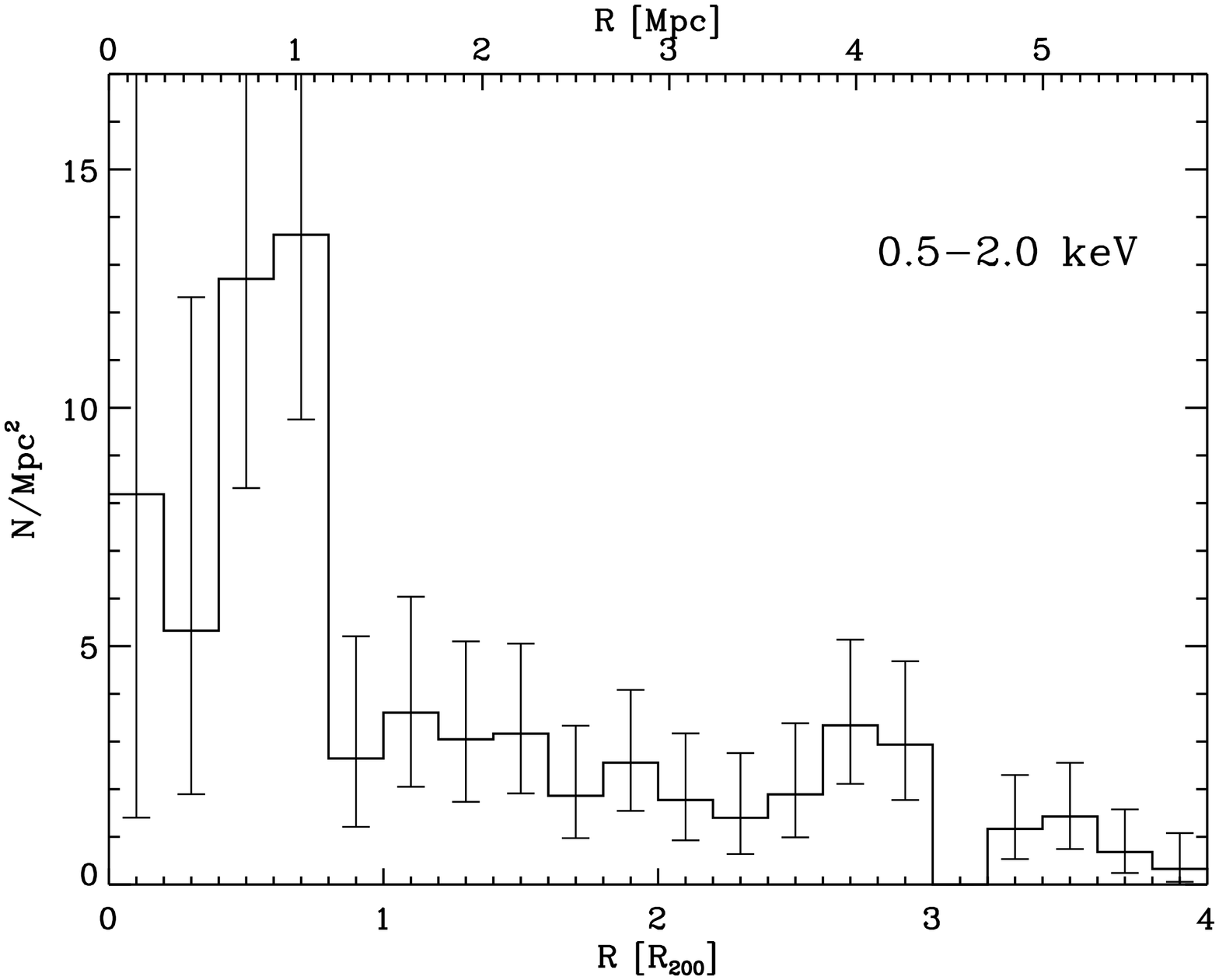}}
\centerline{\includegraphics[angle=90,width=5in]{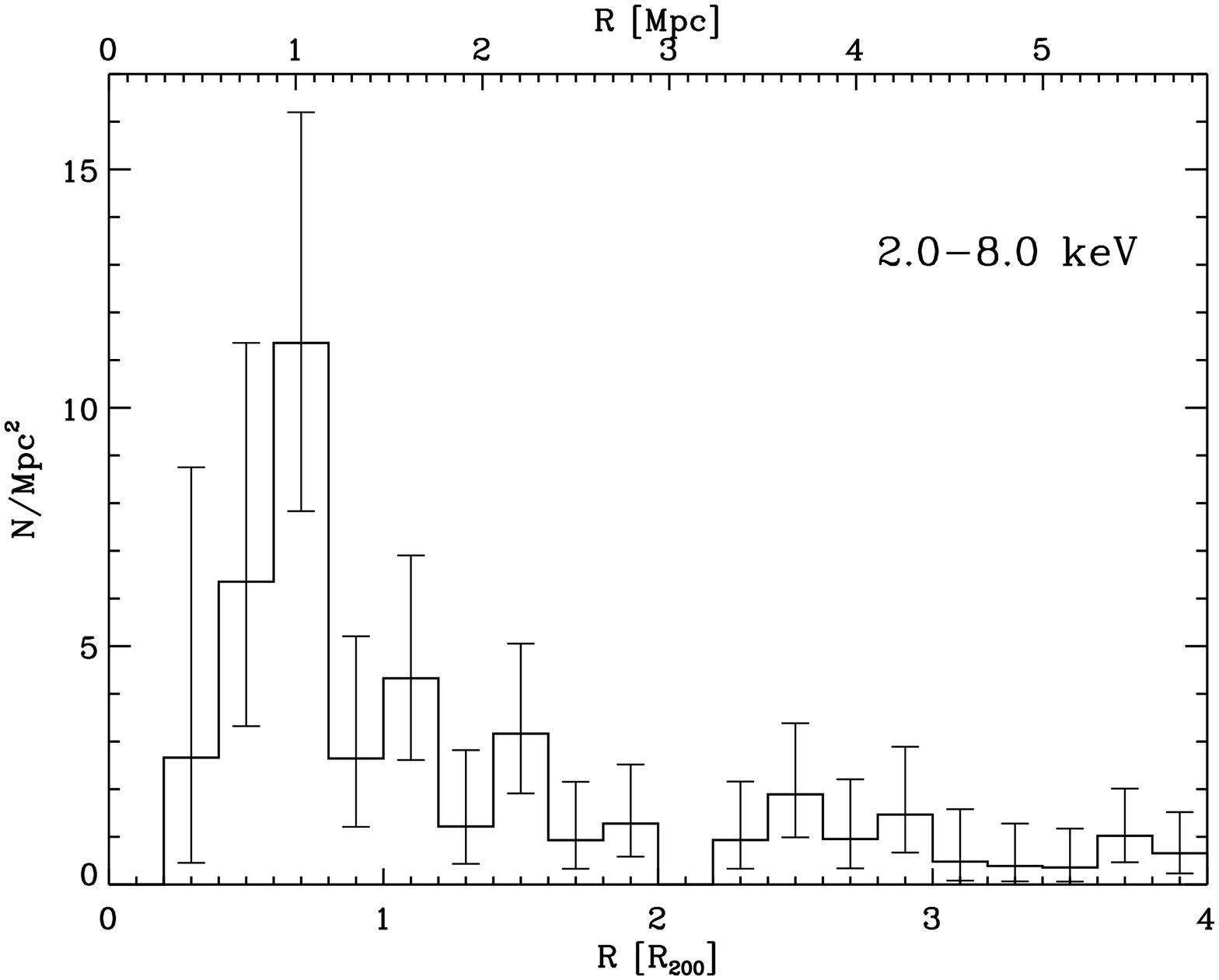}}
\caption{{\bf{Point Source Distribution.}}  Histograms of point source density vs. radius were constructed from data in soft (0.5-2.0 keV) and hard (2-8 keV) energy bands, and are shown here with $1\sigma$ error bars.  Both energy bands indicate an excess of point sources within $~\rm{R}_{200}$.  Each of these plots are consistent with the findings of~\citet{ruderman}.\label{fig8}}
\end{figure}

\begin{figure}
\centerline{\includegraphics[width=4.in]{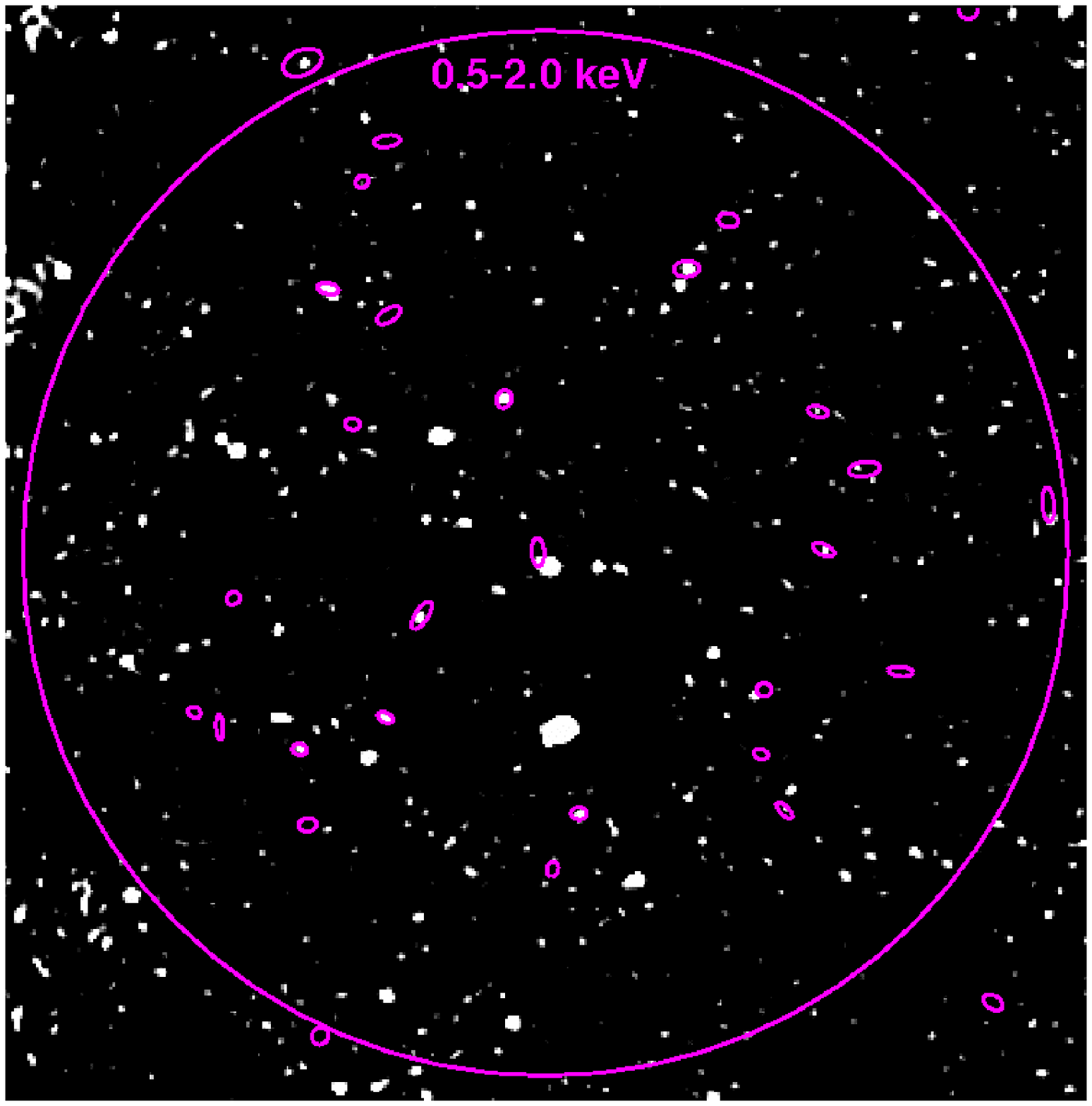}}

\centerline{\includegraphics[width=4.in]{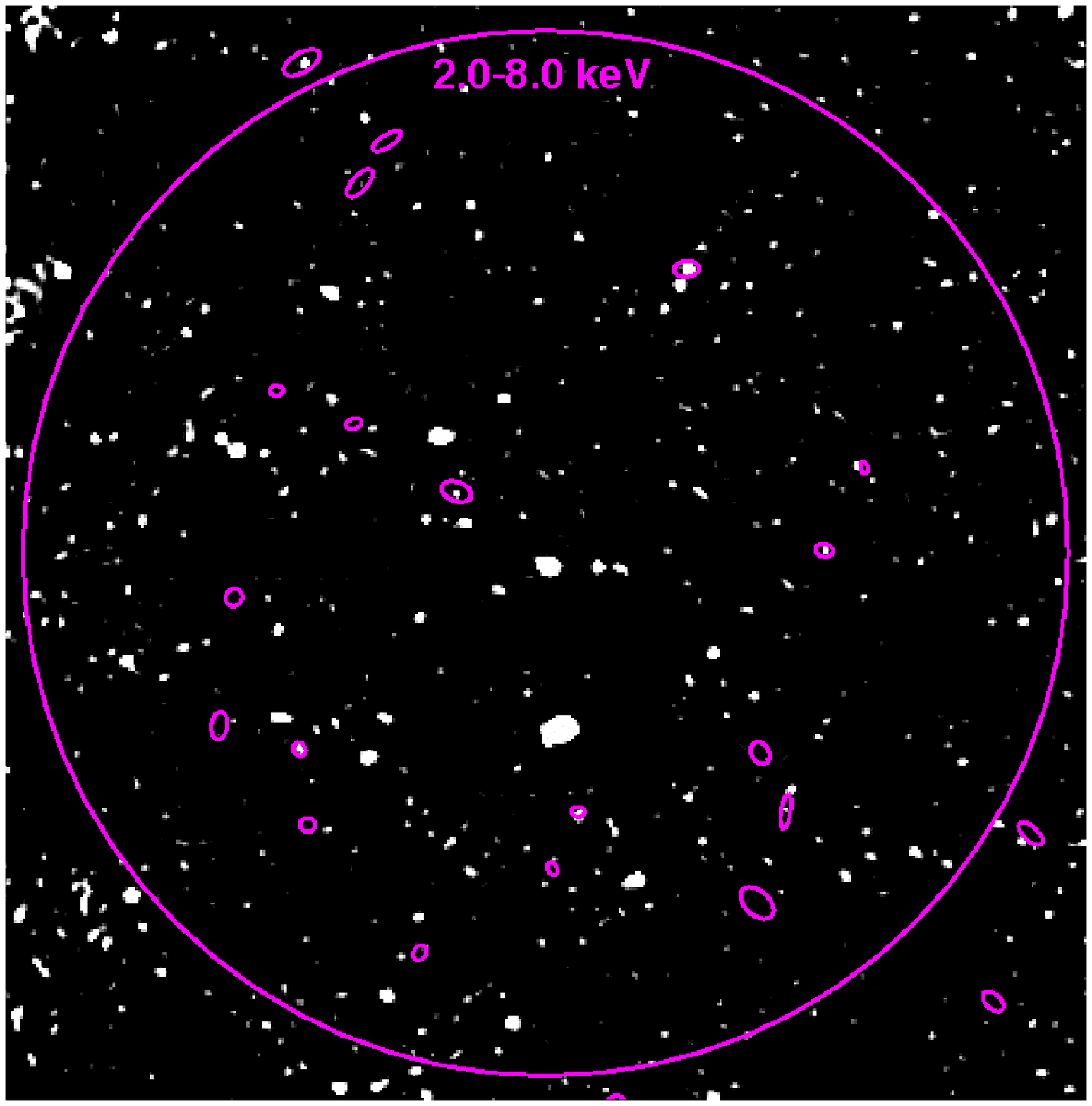}}
\caption{{\bf{VLA Point Source Comparisons.}}  X-ray point source region files produced by running WAVDETECT on 0.5-2.0 keV and 2.0-8.0 keV flux images are overlaid on a 1.4 GHz VLA image of RCS0224-0002.  The size of the point source regions corresponds to the apparent extent of the sources in the X-ray image.  The large circles have a $1.47~h_{70}^{-1}{\rm{Mpc}}~(\rm{R}_{200})$ radius.  Within that region, at least ten of 26 soft band X-ray point sources are found to be coincident with radio emission, and six of 18 detected in the hard X-ray band.  North is up in both images.\label{fig9}}
\end{figure}

Hardness ratio maps were created from soft (0.5-2.0 keV) and hard band (2.0-8.0 keV) flux images.  A peak in soft X-ray emission is seen to coincide with the central peak in surface brightness.  This peak is elongated toward a 1.4 GHz VLA radio source which is centered on the BCG of this cluster, as seen in the HST image (Figure~\ref{fig4}).  


Using the results of both spectral and surface brightness fitting, we have calculated a central density for RCS0224-0002 of $7.6\times10^{-3}\pm{9\times10^{-4}}~\rm{cm}^{-3}$, which results in a central cooling time of $3.3\times 10^9$ years.  Without spatially resolved temperature information we are unable to confirm the presence of a cool component in the center of RCS0224-0002, however the central surface brightness excess, corresponding soft peak of emission, and lack of a central hard point source support the idea that RCS0224-0002 may contain a small proto-cooling core.  Similar evidence has been detected in another $z\sim 0.8$ cluster, MS1137.5+6625~\citep{grego04}.  These two clusters are the highest redshift systems in which central cooling has yet been detected (albeit indirectly), and their existence provides evidence that massive clusters are in place at early times \citep{mullis05}.

Mass estimates of RCS0224-0002 result in total masses of $\rm{M}_{\rm{2500,tot}}=1.6\pm{0.2} \times 10^{14}~h_{70}^{-1} \msun$ and a gas mass of $\rm{M}_{\rm{2500,gas}}=0.050\pm{0.005} \times 10^{14}~h_{70}^{-5/2} \msun$.  This estimate of total gravitating mass is consistent within 68\% errors with weak lensing mass estimates of $M_{2500}=1.3^{+1.0}_{-0.6}\times
10^{14}\msun$.  Similarly, we find a rest-frame velocity dispersion for this cluster of $900\pm180~\rm{km}~{\rm{s}}^{-1}$, which is consistent with that predicted by the $\sigma-T_X$ relationship~\citep{voit03,xue00}, as well as both weak (\ref{s:lens}) and strong lensing estimates of the velocity dispersion~\citep{swinbank07}.  The consistency of these mass estimators is an indication that RCS0224-0002 is a relatively relaxed system.  

An investigation of the point source population in the field of RCS0224-0002 indicates a possible slight excess of low-flux soft energy sources within $\rm{R}_{200}$ of the cluster center.  The high energy source population of this cluster is generally consistent with the measured X-ray background.  Both measurements of point source density represent lower limits because of a selection effect which makes it more difficult to detect point sources close to the cluster center (due to the surface brightness of the cluster).  Radial distributions of these sources exhibit a significant peak within~$\rm{R}_{200}$ in both the hard and soft bands, possibly indicating that the cluster potential is influencing AGN activity at that radius.  Spectral information on cluster member galaxies is needed, however, to test that hypothesis.  The radial profile of soft X-ray point sources is consistent with the results of~\citet{ruderman}.  X-ray point sources within the field of RCS0224-0002 have been compared to 1.4 GHz VLA flux information, and coincident sources are listed in Tables~\ref{table3} and~\ref{table4}.

The core gas mass fraction of this cluster, $f_{g,2500}=0.031\pm{0.005}~h_{70}^{-3/2}$, is more than three times lower than expected universal values~\citep{evrard}.  This is a phenomenon that is also seen in groups, poor clusters, and other high redshift clusters~\citep{dellantonio,sanderson,sadat}, as well as high redshift cluster simulations~\citep{ettori04b}.  It is possible that a comparatively higher fraction of the baryonic mass of this object was converted into stars~\citep{vikhlinin06,nagai07}; however additional analysis is required to confirm or rule out this possibility.  In addition, the AGN component of RCS0224-0002 may provide a significant contribution to the entropy of this system, however it is likely that other non-gravitational processes (i.e. galaxy formation) are also involved.  


%

\acknowledgements  

Support for this work was provided by the National Aeronautics and Space Administration through Chandra Award Numbers GO2-3158X and GO4-5153X issued by the Chandra X-ray Observatory Center, which is operated by the Smithsonian Astrophysical Observatory for and on behalf of the National Aeronautics Space Administration under contract NAS8-03060.  Erica Ellingson also acknowledges support from NSF grant AST-0206154.

A portion of this work was based on observations made with the NASA/ESA Hubble Space Telescope, obtained at the Space Telescope Science Institute, which is operated by the Association of Universities for Research in Astronomy, Inc., under NASA contract NAS 5-26555. These observations are associated with program 9135." 

In addition, some of the data presented herein were obtained at the W.M. Keck Observatory from telescope time allocated to the National Aeronautics and Space Administration through the agency's scientific partnership with the California Institute of Technology and the University of California. The Observatory was made possible by the generous financial support of the W.M. Keck Foundation.

Finally, this paper was also based on observations obtained at the Gemini Observatory, which is operated by the Association of Universities for Research in Astronomy, Inc., under a cooperative agreement with the NSF on behalf of the Gemini partnership: the National Science Foundation (United States), the Particle Physics and Astronomy Research Council (United Kingdom), the National Research Council (Canada), CONICYT (Chile), the Australian Research Council (Australia), CNPq (Brazil) and CONICET (Argentina).

We would also like to thank Phil Armitage, Webster Cash, John Houck, Richard Mushotzky, Craig Sarazin, and Michael Wise for their contributions and input.  













\clearpage

\clearpage

\clearpage

\clearpage

\clearpage

\clearpage

\clearpage

\clearpage

\clearpage

\clearpage


\begin{deluxetable}{crrrrrrrr}\rotate
\tablecolumns{9}
\tablewidth{0pc}
\tablecaption{$\beta$-Model Fits\label{table1}}
\tablehead{ 
\colhead{Bin Size} &\colhead{$R_{\rm c1}$} [$h_{70}^{-1} \rm{kpc}$] &
\colhead{$\beta_1$}  & \colhead{$I_{\rm 01}$\tablenotemark{a}} &
\colhead{$R_{\rm c2}$} [$h_{70}^{-1} \rm{kpc}$] & \colhead{$\beta_2$}  &
\colhead{$I_{\rm 02}$\tablenotemark{a}} & \colhead{$I_{\rm B}$\tablenotemark{a}}  & \colhead{$\chi^2/\rm{DOF}$}}
\startdata 
1\arcsec & $228_{-14}^{+10}$ & $0.936_{- 0.058}^{+ 0.069}$ & $ 6.7_{-0.3}^{+0.2}$ & \nodata & \nodata & \nodata & $2.409_{-0.007}^{+0.007}$ & 223.6/191  \\ 
{} & $283_{-17}^{+17}$ & $1.116_{- 0.069}^{+ 0.056}$ & $5.8_{-0.2}^{+0.2}$ & $1.16_{-0.07}^{+0.09}$ & $0.570_{- 0.035}^{+ 0.042}$ & $34_{-2}^{+3}$ & $2.411_{-0.007}^{+0.007}$ & 218.5/188 \\
2\arcsec & $213_{-13}^{+14}$ & $0.884_{- 0.054}^{+ 0.066}$ & $ 7.1_{-0.4}^{+0.5}$ &  \nodata & \nodata & \nodata & $2.414_{-0.01}^{+0.01}$ & 153.8/130  \\
{} & $267_{-16}^{+16}$ & $1.038_{- 0.064}^{+ 0.073}$ & $ 6.0_{-0.2}^{+0.4}$ & $3.49_{-0.21}^{+0.26}$ & $0.620_{- 0.038}^{+ 0.046}$ & $ 6.3_{-0.4}^{+0.5}$ & $2.416_{-0.01}^{+0.01}$ & 151.5/127  \\

\enddata
\tablenotetext{a}{Surface brightness $I$ in units of $10^{-9}$ photons sec${}^{-1}$ cm${}^{-2}$ arcsec${}^{-2}$}
\end{deluxetable}

\clearpage

\begin{deluxetable}{ccccc}
\tabletypesize{\small}
\tablecolumns{5}
\tablewidth{0pt}
\tablecaption{Summary of the optical multi-object spectroscopic observations.\label{table2}}
\tablehead{
\colhead{Telescope ¥} & \colhead{Instrument} & \colhead{Date} & \colhead{Spectral Range} &  \colhead{Resolution}  \\
}
\startdata
    Keck &  LRIS & 11-21-2001 & 5800-8000 \AA  &  7.5 \AA   \\
    VLT & FORS-2 & 01-27-2003& 4500-8500 \AA &  15 \AA \\ 
    & & 01-28-2003 \\
      & & 01-30-2003 \\
    Gemini & GMOS-N & 10-21-2003 & 4600-9000 \AA & 16 \AA \\ 
     &  & 10-22-2003\\
     &  & 10-23-2003\\ 
  \enddata
\end{deluxetable}     

\clearpage

\begin{deluxetable}{cccccccc}
\tabletypesize{\small}
\tablecolumns{8}
\tablewidth{0pt}
\tablecaption{0.5-2.0 keV Point Sources for ${\rm{R}}~<~{\rm{R}}_{200}$\label{table3}}
\tablehead{
\colhead{Number\tablenotemark{a}} & 
\colhead{RA} & 
\colhead{Dec} &
\colhead{Flux\tablenotemark{b}} &
\colhead{${\rm{L}}_{\rm{x}}$\tablenotemark{c}} &
\colhead{$\sigma$} & 
\colhead{Distance} &
\colhead{1.4 GHz}  \\
\colhead{} & 
\colhead{[J2000]} &
\colhead{[J2000]} &
\colhead{} & 
\colhead{[$10^{42}~{\rm{erg}}~{\rm{s}}^{-1}$]} & 
\colhead{} & 
\colhead{[$h_{70}^{-1}$ Mpc]} &
\colhead{} 
}
\startdata
 1 & 02:24:34.344 & -00:02:26.05 & $3.5\pm{1.3}$ & $0.61\pm{0.23}$ & 3.4 & 0.022 & \nodata\\
 2 & 02:24:37.272 & -00:02:49.89 & $6.3\pm{1.8}$ & $1.11\pm{0.31}$ & 5.7 & 0.395 & yes\\
 3\tablenotemark{d} & 02:24:35.208 & -00:01:28.09 & $7.9\pm{1.9}$ & $1.39\pm{0.34}$ & 7.7 & 0.456 & yes\\
 4 & 02:24:38.208 & -00:03:28.67 & $3.2\pm{1.2}$ & $0.56\pm{0.22}$ & 3.3 & 0.655 & yes\\
 5 (2)\tablenotemark{d} & 02:24:39.024 & -00:01:37.62 & $30.2\pm{3.7}$ & $5.30\pm{0.65}$ & 27.3 & 0.662 & \nodata\\
 6 & 02:24:28.656 & -00:03:18.10 & $9.9\pm{2.1}$ & $1.75\pm{0.38}$ & 9.5 & 0.735 & \nodata\\
 7 (3)& 02:24:33.312 & -00:04:04.98 & $5.8\pm{1.6}$ & $1.02\pm{0.29}$ & 5.9 & 0.749 & yes\tablenotemark{e}\\
 8 (4)& 02:24:27.144 & -00:02:25.11 & $6.1\pm{1.8}$ & $1.07\pm{0.32}$ & 5.4 & 0.794 & yes\\
 9 & 02:24:38.112 & -00:00:56.40 & $3.2\pm{1.3}$ & $0.56\pm{0.22}$ & 3.1 & 0.813 & \nodata\\
10 (5)\tablenotemark{d}& 02:24:28.728 & -00:03:42.58 & $32.8\pm{3.8}$ & $5.76\pm{0.67}$ & 28.2 & 0.841 & \nodata\\
11 & 02:24:27.288 & -00:01:32.81 & $5.7\pm{1.7}$ & $1.00\pm{0.29}$ & 5.3 & 0.876 & yes\\
12 (8)\tablenotemark{d}& 02:24:40.368 & -00:03:40.72 & $61.3\pm{5.2}$ & $10.78\pm{0.92}$ & 56.5 & 0.898 & yes\\
13 (7)& 02:24:42.048 & -00:02:43.56 & $18.5\pm{2.9}$ & $3.25\pm{0.51}$ & 17.9 & 0.901 & \nodata\\
14 (9)& 02:24:33.984 & -00:04:26.01 & $24.8\pm{3.3}$ & $4.36\pm{0.58}$ & 23.7 & 0.902 & \nodata\\
15 (10)& 02:24:30.600 & -00:00:38.78 & $1429.4\pm{25.3}$ & $251.23\pm{4.44}$ & 916.0 & 0.906 & yes\\
16 (11)& 02:24:26.112 & -00:01:54.61 & $8.0\pm{1.9}$ & $1.40\pm{0.34}$ & 7.6 & 0.941 & ?\\
17 & 02:24:39.648 & -00:00:46.28 & $5.3\pm{1.6}$ & $0.93\pm{0.28}$ & 4.8 & 0.977 & yes\\
18 (12)& 02:24:28.128 & -00:04:03.81 & $5.3\pm{1.6}$ & $0.93\pm{0.28}$ & 5.4 & 1.002 & yes\\
19 (13)& 02:24:40.176 & -00:04:09.33 & $39.5\pm{4.2}$ & $6.95\pm{0.74}$ & 36.9 & 1.031 & \nodata\\
20 (14)& 02:24:42.384 & -00:03:32.32 & $4.5\pm{1.5}$ & $0.79\pm{0.26}$ & 4.5 & 1.055 & \nodata\\
21 & 02:24:25.200 & -00:03:11.21 & $5.4\pm{1.6}$ & $0.94\pm{0.28}$ & 5.1 & 1.069 & \nodata\\
22 & 02:24:29.544 & -00:00:20.37 & $29.2\pm{3.7}$ & $5.13\pm{0.64}$ & 25.2 & 1.084 & \nodata\\
23 & 02:24:43.032 & -00:03:26.71 & $7.6\pm{1.9}$ & $1.34\pm{0.33}$ & 7.7 & 1.101 & \nodata\\
24 (16)& 02:24:38.784 & -00:00:05.81 & $8.0\pm{2.0}$ & $1.41\pm{0.36}$ & 6.9 & 1.183 & ?\\
25 (18)& 02:24:38.160 & +00:00:09.60 & $5.6\pm{1.6}$ & $0.98\pm{0.29}$ & 5.4 & 1.260 & \nodata\\
26\tablenotemark{d} & 02:24:21.456 & -00:02:08.01 & $3.6\pm{1.4}$ & $0.64\pm{0.24}$ & 3.4 & 1.442 & \nodata\\

\enddata

\tablenotetext{a}{0.5-2.0 keV (2.0-8.0 keV)} 
\tablenotetext{b}{Unabsorbed 0.5-2.0 keV flux in units of $10^-16~\rm{erg}~\rm{cm}^{-2}~\rm{s}^{-1}$} 
\tablenotetext{c}{Unabsorbed bolometric X-ray luminosity obtained using 0.5-2.0 keV count rates and PIMMS, assuming z=0.778 and a powerlaw index of 1.6} 
\tablenotetext{d}{Within $1\arcsec$ of a possible cluster member, using photometrically determined redshifts}
\tablenotetext{e}{SCUBA submillimeter source \citep{webb}}
\end{deluxetable}

\clearpage

\begin{deluxetable}{cccccccc}
\tabletypesize{\small}
\tablecolumns{8}
\tablewidth{0pt}
\tablecaption{2.0-8.0 keV Point Sources for $\rm{R}~<~\rm{R}_{200}$\label{table4}}
\tablehead{
\colhead{Number\tablenotemark{a}} & 
\colhead{RA} & 
\colhead{Dec} &
\colhead{Flux\tablenotemark{b}} &
\colhead{${\rm{L}}_{\rm{x}}$\tablenotemark{c}} &
\colhead{$\sigma$} & 
\colhead{Distance}  &
\colhead{1.4 GHz} \\
\colhead{} & 
\colhead{[J2000]} &
\colhead{[J2000]} &
\colhead{} & 
\colhead{[$10^{42}~{\rm{erg}}~{\rm{s}}^{-1}$]} & 
\colhead{} & 
\colhead{[$h_{70}^{-1}$ Mpc]} &
\colhead{} 
}
\startdata
 1 & 02:24:36.408 & -00:02:03.15 & $19.4\pm{6.9}$ & $1.96\pm{0.70}$ & 3.5 & 0.309 & yes\\
 2 (5)\tablenotemark{d}& 02:24:39.000 & -00:01:37.52 & $34.3\pm{8.8}$ & $3.46\pm{0.89}$ & 6.4 & 0.660 & \nodata\\
 3 (7)& 02:24:33.336 & -00:04:04.34 & $36.9\pm{8.9}$ & $3.72\pm{0.90}$ & 7.3 & 0.744 & yes\tablenotemark{e}\\
 4 (8)& 02:24:27.120 & -00:02:25.40 & $107.5\pm{16.0}$ & $10.85\pm{1.62}$ & 16.1 & 0.796 & yes\\
 5 (10)\tablenotemark{d}& 02:24:28.728 & -00:03:42.14 & $28.0\pm{8.0}$ & $2.83\pm{0.81}$ & 5.3 & 0.838 & \nodata\\
 6\tablenotemark{d} & 02:24:40.944 & -00:01:24.94 & $19.5\pm{6.8}$ & $1.97\pm{0.69}$ & 3.8 & 0.897 & \nodata\\
 7 (13)& 02:24:42.024 & -00:02:43.26 & $36.1\pm{9.0}$ & $3.64\pm{0.91}$ & 6.7 & 0.898 & \nodata\\
 8 (12)\tablenotemark{d}& 02:24:40.368 & -00:03:40.75 & $119.8\pm{15.9}$ & $12.10\pm{1.61}$ & 19.9 & 0.899 & yes\\
 9 (14)& 02:24:33.984 & -00:04:25.94 & $37.0\pm{8.9}$ & $3.73\pm{0.90}$ & 7.5 & 0.901 & \nodata\\
10 (15)& 02:24:30.600 & -00:00:38.86 & $862.3\pm{42.2}$ & $87.05\pm{4.26}$ & 119.0 & 0.906 & yes\\
11 (16)& 02:24:26.112 & -00:01:54.13 & $16.1\pm{6.1}$ & $1.63\pm{0.62}$ & 3.4 & 0.942 & \nodata\\
12 (18)& 02:24:28.104 & -00:04:04.18 & $20.3\pm{6.9}$ & $2.05\pm{0.70}$ & 3.9 & 1.007 & yes\\
13 (19)& 02:24:40.176 & -00:04:09.44 & $74.1\pm{12.6}$ & $7.48\pm{1.27}$ & 13.6 & 1.032 & \nodata\\
14 (20)\tablenotemark{d}& 02:24:42.384 & -00:03:31.73 & $37.4\pm{9.2}$ & $3.77\pm{0.93}$ & 6.6 & 1.053 & \nodata\\
15 & 02:24:28.824 & -00:04:38.94 & $20.0\pm{7.3}$ & $2.02\pm{0.73}$ & 3.4 & 1.167 & \nodata \\
16 (24)& 02:24:38.856 & -00:00:06.35 & $27.5\pm{8.4}$ & $2.78\pm{0.85}$ & 4.6 & 1.182 & \nodata\\
17\tablenotemark{d} & 02:24:37.344 & -00:04:57.77 & $124.6\pm{16.3}$ & $12.58\pm{1.64}$ & 20.4 & 1.196 & \nodata\\
18 (25)& 02:24:38.160 & +00:00:09.53 & $74.0\pm{12.9}$ & $7.47\pm{1.30}$ & 12.6 & 1.260 & \nodata\\
\enddata

\tablenotetext{a}{2.0-8.0 keV (0.5-2.0 keV)}
\tablenotetext{b}{Unabsorbed 2.0-8.0 keV flux in units of $10^-16~\rm{erg}~\rm{cm}^{-2}~\rm{s}^{-1}$} 
\tablenotetext{c}{Unabsorbed bolometric X-ray luminosity obtained using 2.0-8.0 keV count rates and PIMMS, assuming z=0.778 and a powerlaw index of 1.6}  
\tablenotetext{d}{Within $1\arcsec$ of a possible cluster member, using photometrically determined redshifts}
\tablenotetext{e}{SCUBA submillimeter source \citep{webb}}
\end{deluxetable}




\end{document}